\documentclass[a4paper,12pt]{article}%
\usepackage[latin1]{inputenc}
\usepackage{amsfonts}
\usepackage{amsmath}
\usepackage{amssymb}
\usepackage{authblk}
\usepackage{multirow}
\usepackage{hyperref}
\usepackage{graphicx,color}
\usepackage[onehalfspacing]{setspace}
\newtheorem{theorem}{Theorem}
\newtheorem{corollary}{Corollary}
\newtheorem{lemma}{Lemma}
\newtheorem{remark}{Remark}
\newcommand{\bfmath}[1]{\mbox{\boldmath$#1$\unboldmath}}

\title{\textbf{Optimal estimation in functional linear regression for sparse noise-contaminated data}}
\author[1]{Behdad Mostafaiy \thanks{{\scriptsize Department of Statistics, University of Mohaghegh Ardabili, Ardabil, Iran}}}
\affil[1]{University of Mohaghegh Ardabili}
\author[2]{MohammadReza FaridRohani \thanks{{\scriptsize Department of Statistics, Shahid Beheshti University, Tehran, Iran 19834}} }
\affil[2]{Shahid Beheshti University}

\author[3]{\\ Shojaeddin Chenouri \thanks{{\scriptsize Department of Statistics and Actuarial Science, University of Waterloo, Waterloo, ON, Canada.}}\thanks{{\scriptsize Corresponding author: schenouri@uwaterloo.ca}}}
\affil[3]{University of Waterloo}

\date{
  }
\begin{document}

\maketitle

\pagestyle{myheadings}

\begin{abstract}
In this paper, we propose a novel approach to fit a functional linear regression in which both the response and the predictor are functions of a common variable such as time. We consider the case that the response and the predictor processes are both sparsely sampled on random time points and are contaminated with random errors. In addition, the random times are allowed to be different for the measurements of the predictor and the response functions. The aforementioned situation often occurs in the longitudinal data settings. To estimate the covariance and the cross-covariance functions we use a regularization method over a reproducing kernel Hilbert space. The estimate of the cross-covarinace function is used to obtain an estimate of the regression coefficient function and also functional singular components. We derive the convergence rates of the proposed cross-covariance, the regression coefficient and the singular component function estimators. Furthermore, we show that, under some regularity conditions, the estimator of the coefficient function has a minimax optimal rate. We conduct a simulation study and  demonstrate merits of the proposed method by comparing it to some other existing methods in the literature.  We illustrate the method by an example of an application to a well known multicenter AIDS Cohort Study.  
\end{abstract}
\section{Introduction}
Functional data analysis is concerned with experiments in which each observation is a curve or a $d$-dimensional surface. This is an extension of multivariate data analysis when observations are replaced by infinite dimensional vectors rather than finite. There are many techniques available for analysis of functional data in the literature. Ramsay and Silverman (2002, 2005) provided an overview of applications and available techniques up-to that date. See also Ferraty and Vieu (2006), and Ramsay {\it et al.} (2009), Horv\'{a}th and Kokoszka (2012), Hsing and Eubank (2015).  

Functional linear regression refers to a class of problems that a response is related to one or more predictors which either response or some of the predictors are functions. The basic idea of functional regression  appeared long ago in Grenander (1950), but it became popular after the work of Ramsay and Dalzell (1991), where they proposed a penalized least squares method to estimate the functional linear regression coefficient surface. Since then an extensive amount of work has been done. See for example M\"{u}ller and Stadtm\"{u}ller (2005), Ramsay and Silverman (2005), Cai and Hall (2006), Cardot and Sarda (2006), Hall and Horowitz (2007), Li and Hsing (2007), Preda (2007), Shin (2009), Yuan and Cai (2010),  Cai and Yuan (2012), Shin and Hsing (2012), and Shin and Lee (2012) for the case when the response is a scaler or vector. In the case that both response and predictors are functions, see Ramsay and Silverman (2005) or Horv\'{a}th and Kokoszka (2012), for  summaries of techniques and applications, He {\it et al.} (2000), M\"{u}ller (2005), Yao {\it et al.} (2005b), and Ferraty {\it et al.} (2012), Ivanescu {\it et al.} (2015).  
 
Most approaches to functional linear regression are based on functional principle component analysis (FPCA). Examples are Cardot and Sarda (2003), Yao \textit{et al.,} (2005b), Cai and Hall (2006), Hall and Horowitz (2007). Some methods are based on regularization in some suitable spaces, for example, Li and Hsing (2007), Preda (2007), Yuan and Cai (2010) and Cai and Yuan (2012). Covariance and cross-covariance functions of the involved response and predictors have crucial roles in these models. Since the covariance and cross-covariance functions are typically unknown in practice, it is important to introduce some suitable methods to estimate these functions. 

Majority of the aforementioned papers are developed for experiments involving functional data that are generated on densely sampled grids. In many experiments though, for example most longitudinal studies, the functional trajectories of the involved smooth random processes are not densely and directly observable. In these cases, the observed data are noisy, sparse and irregularly spaced measurements of these trajectories. By sparseness, we mean that the sampling frequency of the curves are relatively small as in Cai and Yuan (2011). 

Following the notation in Yao {\it et al.} (2005a), let $U_{i\,j}$ and $V_{i\,j}$ be the $j$th observations of the random trajectories $X_i(\cdot)$ and $Y_i(\cdot)$ at a random time points $S_{i\,j}$ and $T_{i\,j}$, respectively. Assume that $S_{i\,j}$ and $T_{i\,j}$ are independently drawn from a common distribution on a compact domain $\mathcal{T}\subset \mathbb{R}$. In addition, assume that $U_{i\,j}$ and $V_{i\,j}$ are contaminated with measurement errors $\varepsilon_{i\,j}$ and  $\epsilon_{i\,j}$, respectively. These errors are assumed to be $i.i.d.$ with mean zero and finite variance $\sigma^2_{_X}$ for $\varepsilon_{i\,j}$ and $\sigma^2_{_Y}$ for $\epsilon_{i\,j}$. Therefore, the models may be represented in the following forms:
\begin{equation}\label{UV}
\begin{split}
U_{ij}&=X_i(S_{ij})+\varepsilon _{ij}, \qquad j=1,\,\dots,\, m_1; \qquad i=1,\,\dots,\,n\,,\\
V_{ij}&=Y_i(T_{ij})+\epsilon _{ij}, \qquad j=1,\,\dots,\, m_2; \qquad i=1,\,\dots,\,n\,.
\end{split}
\end{equation}
To have a better illustration of the methodology we assume, without loss of generality, that $m_1$ and $m_2$ are fixed and equal for all respective curves. Some discussion is presented in Remark \ref{RM5} in Section \ref{S4}.

Recently, Cai and Yuan (2010) studied the covariance function estimation of $X$ in the functional data framework of
$$U_{ij}=X_i(S_{ij})+\varepsilon _{ij}, \qquad j=1,\,\dots,\, m_1; \qquad i=1,\,\dots,\,n\,.$$
Cai and Yuan (2010) utilized a regularization method applying a reproducing kernel Hilbert space (RKHS) framework, and showed that if the sample path of the involved process $X$ belongs to a certain RKHS then the covariance function belongs to a tensor product space. 

In this paper, we consider functional linear regression models of the form  
\begin{equation}\label{flr1.1}
Y(t) =\alpha (t)+ \int_\mathcal{T} X(s)\,\beta (s,\,t)\,ds  + \eta (t)\,
\end{equation}
with measurements through model \eqref{UV}, where $\beta $ is an unknown square integrable coefficient function, $\alpha$ is an intercept function and $\eta$ is a random noise term with zero mean and finite variance. In this model, which is generally known as functional linear model (FLM), the value of $Y(t)$, at any given time point $t$, depends on the entire trajectory of the process $X$, and the coefficient function $\beta(s,\,t)$ can be interpreted as the relative weight placed on $X(s)$ at time $s$ which is required to predict $Y(t)$ on a fixed time $t$.

He \textit{et al.,} (2000) provided a basis representation of the coefficient function $\beta$ in \eqref{flr1.1}, and discussed some of its properties. Yao \textit{et al.,} (2005b) further discussed this model, and provided a method of estimating the regression coefficient function by utilizing the basis  representation of $\beta$ and estimating the involved parameters in this representation using local linear smoothers. Ivanescu {\it et al.} (2015) used an expansion of the smooth coefficient function $\beta$ based on a set of pre-determined bases, and applied a quadratic roughness penalty. The smoothing parameter in the corresponding mixed model representation of the penalized regression in Ivanescu {\it et al.} (2015) was then estimated using the restricted maximum likelihood (REML). 
In the present paper, we take a different approach, which is inspired by the regularization method discussed in Cai and Yuan (2010). We first obtain a regularized estimates of the cross-covariance function of $X$ and $Y$, as well as the marginal covariance functions of $X$ and $Y$, and then use the representation in He \textit{et al.} (2000) to estimate the coefficient function $\beta$. To study the quality of the estimators, we derive the convergence rate of the proposed estimator of the cross-covariance function in $L_2$-sense and show that it has similar properties of the covariance function estimator outlined in Cai and Yuan (2010). In addition, we obtain the rate of convergence of the proposed estimator of the coefficient function $\beta$ under the integrated squared error loss, when assuming sample paths of $X$ and $Y$ are differentiable of certain order. Under some regularity conditions, we also show that the proposed estimator of the coefficient function $\beta$ has an optimal rate in minimax sense.  

Functional singular component analysis (FSCA) is another subject that we are pursuing in this paper. Yang \textit{et al.,} (2011) developed the concept of functional singular value decomposition for covariance functions and functional singular component analysis. They introduced a method of estimation for singular values and functions for the case of sparse and noise-contaminated longitudinal data. They also discussed the asymptotic properties of these estimators. In this paper, we use our proposed cross-covariance function estimator to derive estimators of singular values and singular functions of $X$ and $Y$ in closed forms making them computationally very tractable. In addition, we provide rates of convergence in $L_2$-sense for these singular values and functions estimators.

In summary, the main contributions of this paper are the followings:
\begin{itemize}
\item[a) ] We employ a regularization method on an RKHS to estimate the cross-covariance function of the processes $X$ and $Y$, when both processes are sparsely and irregularly observed, and observations are contaminated with measurement errors.  We derive the rate of convergence of the proposed estimator under the integrated squared error loss.    
\item[b) ] We estimate the coefficient function $\beta$ by using the cross-covariance estimate in (a) and derive the optimal rate of convergence for the proposed estimator of $\beta$ under the integrated squared error loss. In addition, we study the finite sample behaviour of this estimator via simulation and show that it produces smaller mean square, and mean absolute errors in compare to the competing methods in the literature.
\item[c) ] We estimate the singular values and singular functions of $X$ and $Y$ by utilizing the cross-covariance estimate in (a), and provide rates of convergence for these estimators. It is noteworthy that these estimates are in closed forms making them computationally very tractable. 
\end{itemize}

Organization of the article is as follows. In Section 2, after reviewing  basic notation, definitions and results, we establish that the cross-covariance functions of $X$ and $Y$ belongs to a tensor product Hilbert space. In Section 3, we utilize a regularization method to estimate the the cross-covariance functions and then provide estimates of the coefficient function $\beta$, and the singular values and functions. In Section 4, we establish rates of convergence for the estimators obtained in Section 3. In Section 5, we present a  simulation study and also apply the proposed method on a real dataset. We give the proofs in Section 6. 
\section{Notation and fundamental concepts and results}
In this section, we provide notation and some basic definitions, and background results to be used throughout the paper. In addition, we establish that the cross-covariance function of $X$ and $Y$, and some other related kernel functions belong to a tensor product Hilbert space.    
\subsection{Spectral analysis of stochastic processes}
Let $X$ and $Y$ be second-order stochastic processes on compact domains. For the sake of simplicity of presentation, without loss of generality, we assume that $X$ and $Y$ have a common domain $\mathcal{T}\subset \mathbb{R}$. For a discussion on this we refer to Remark \ref{RM7} in Section \ref{S4}. Let for $s\in \mathcal{T}$,  $\mu _{_X}(s)=\text{E}[X(s)]$ and $\mu _{_Y}(s)=\text{E}[Y(s)]$ represent the mean functions of $X$ and $Y$, respectively.  
The covariance functions are
$$C_{_{XX}}(s,\,t)=\text{cov}(X(s),X(t)), \quad C_{_{Y\,Y}}(s,\,t)=\text{cov}(Y(s),Y(t))$$
and cross-covariance functions are
$$C_{_{XY}}(s,t)=\text{cov}(X(s),Y(t)), \quad C_{_{YX}}(t,s)=\text{cov}(Y(t),X(s))\,,$$
where 
$$\text{cov}(X(s),Y(t))=\text{E}\left[(X(s)-\mu _{_X}(s))\,(Y(t)-\mu _{_Y}(t)) \right]\qquad \text{for all } s,\,t\in\mathcal{T}\,.$$
The assumption of square integrality of the processes $X$ and $Y$, that is
\begin{equation*}
\text{E}\left[\,\|X\|_{_{L_2 (\mathcal{T})}}^2\, \right] = \text{E}\left[  \int_\mathcal{T}  X^2 (s)\,ds \right] < \infty \,,\quad
\text{E}\left[\,\|Y\|_{_{L_2(\mathcal{T})}}^2\right]=\text{E}\left[  \int_\mathcal{T}  Y^2 (s)\,ds \right]<\infty\,
\end{equation*}
along with twice application of the Cauchy-Schwarz inequality imply that $C_{_{XX}}$ and $C_{_{YY}}$ are also square integrable, that is
\begin{equation}
\int_{\mathcal{T} \times \mathcal{T}} {C_{_{XX}}^2 (s,\,t )\,ds\, dt }  < \infty ,\hspace{.4cm}
\int_{\mathcal{T} \times \mathcal{T}} {C_{_{YY}}^2 (s,\,t)\,ds\, dt }  < \infty .
\end{equation}
Now from the Mercer's theorem, the covariance functions $C_{_{XX}}$ and $C_{_{YY}}$ admit the following spectral expansions:
\begin{equation}\label{CXCY}
C_{_{XX}} (s_1 ,\,s_2 ) = \sum\limits_{i = 1}^\infty \lambda_{_{X \,i}}\, \Psi_i (s_1 )\,\Psi_i (s_2 )\,, \qquad 
C_{_{YY}} (t_1 ,\,t_2 ) = \sum\limits_{j = 1}^\infty \lambda_{_{Y\, j}} \,\Phi_j (t_1 )\,\Phi_j (t_2 )\,,
\end{equation}
where $\lambda _{_{X \,1}}\geq \lambda _{_{X\, 2}}\geq \dots \geq 0$ and $\lambda _{_{Y \,1}}\geq \lambda _{_{Y\, 2}}\geq \dots \geq 0$ are the eigenvalue sequences of integral operators with kernels $C_{_{XX}}$ and $C_{_{YY}}$, respectively, and 
 $\Psi _1,\,\Psi _2,\,\dots$ and $\Phi _1,\,\Phi _2,\,\dots$ are the corresponding orthonormal eigenfunction sequences.
 
\noindent 
From these, we have the Karhunen-Loeve expansions for $X$ and $Y$ in forms of
\begin{equation}\label{KLexp}
X(s) =\mu _{_X}(s)+ \sum\limits_{i = 1}^\infty  \zeta _i\, \Psi _i (s),\qquad
Y(t) =\mu _{_Y}(t)+ \sum\limits_{j = 1}^\infty  \xi _j \,\Phi _j (t)\,,\qquad \text{for } s,\,t\,\in\mathcal{T}\,,
\end{equation}
where $\zeta _i$s and $\xi _j$s are independent random variables with $E[\,\zeta _i\,] =0=E[\,\xi _j\,] =0$ and
 $E[\,\zeta _i ^2\,]=\lambda _{_{X \,i}}$ and $E[\,\xi _j ^2\,]=\lambda _{_{Y\,j}}$, respectively. 

\noindent
Let $\mathcal{C}_{_{X\,Y}}$ and $\mathcal{C}_{_{Y\,X}}$ be the integral operators with corresponding kernels given by the cross-covariance functions $C_{_{X\,Y}}$ and $C_{_{Y\,X}}$, i.e.
\begin{equation*}
\mathcal{C}_{_{X\,Y}} (f)(\cdot) = \int_\mathcal{T} {C_{_{X\,Y}} (\cdot,\,t)\,f(t)\,dt}\,,\qquad
\mathcal{C}_{_{Y\,X}} (f)(\cdot) = \int_\mathcal{T} {C_{_{Y\,X}} (\cdot,\,t)\,f(t)\,dt}\,.
\end{equation*}
Define $\mathcal{A}_{_{X\,Y}}=\mathcal{C}_{_{X\,Y}}\,\circ\, \mathcal{C}_{_{Y\,X}}$ and  $\mathcal{A}_{_{Y\,X}}=\mathcal{C}_{_{Y\,X}}\,\circ \,\mathcal{C}_{_{X\,Y}}$. Because $\mathcal{C}_{_{X\,Y}}$ is the adjoint operator of $\mathcal{C}_{_{Y\,X}}$, then $\mathcal{A}_{_{X\,Y}}$ and $\mathcal{A}_{_{Y\,X}}$ are self-adjoint and Hilbert-Schmidt operators with $L_2(\mathcal{T}\times \mathcal{T})$-kernels  
\begin{equation*}
A_{_{X\,Y}} (s,\,t) = \int_\mathcal{T} C_{_{X\,Y}} (s,\,u)\,C_{_{X\,Y}} (t,\,u)\,du\,,\quad
A_{_{Y\,X}} (s,\,t) = \int_\mathcal{T} C_{_{X\,Y}} (u,\,s)\,C_{_{X\,Y}} (u,\,t)\,du\,.
\end{equation*}
On the other hand $\mathcal{A}_{_{X\,Y}}$ and $\mathcal{A}_{_{Y\,X}}$ have common eigenvalues $\sigma _1^2  \geq \sigma _2^2  \geq\dots \geq 0$ and orthonormal eigenfunctions $\phi _1,\,\phi _2,\,\dots$ for $\mathcal{A}_{_{XY}}$ and $\psi _1,\,\psi _2,\,\dots$ for $\mathcal{A}_{_{YX}}$, respectively. See Kato (1995), Chapter V, Section 3, pages 260-266 for more details. In addition, we have $\mathcal{C}_{_{XY}}(\phi _k)=\sigma _k \,\psi _k$ and $\mathcal{C}_{_{YX}}(\psi _k)=\sigma _k\, \phi _k$. The sequences $\phi _1,\,\phi _2,\,\dots$ and $\psi _1,\,\psi _2,\,\dots$ are called singular functions and the sequence $\sigma _1,\,\sigma _2,\,\dots$ is called singular values. See Yang \textit{et al.} (2011) and references therein.

We can expand the integral operators $\mathcal{C}_{_{X\,Y}}$ and $\mathcal{C}_{_{Y\,X}}$ as
\begin{equation*}
\mathcal{C}_{_{X\,Y}}(f)(\cdot) = \sum\limits_{k = 1}^\infty  \sigma _k\, \langle\, f,\,\psi _k \,\rangle \,\phi _k(\cdot),\qquad
\mathcal{C}_{_{Y\,X}}(f)(\cdot) = \sum\limits_{k = 1}^\infty  \sigma _k\, \langle \,f,\,\phi _k\, \rangle \,\psi _k(\cdot)\,,
\end{equation*}
and these imply the expansions
\begin{equation*}
C_{_{X\,Y}}(s,\,t) = \sum\limits_{k = 1}^\infty \sigma _k\,\psi _k(s)\,\phi _k(t)\quad \text{ and }\quad 
C_{_{Y\,X}}(s,\,t) = \sum\limits_{k = 1}^\infty  \sigma _k\,\phi _k(s)\,\psi _k(t)\,.
\end{equation*}
\subsection{Reproducing Kernel Hilbert Spaces}
The theory of Reproducing Kernel Hilbert Spaces (RKHS) plays an important rule in this paper, so it is helpful to review some of the basic facts about them. More details can be found in Aronszajn (1950), Wahba (1990), Berlinet and Thomas-Agnan (2004), and Hsing and Eubank (2015). 
A Hilbert space $\mathcal{H}$ of functions on a set $\mathcal{T}$ with inner product $\langle\cdot\,,\,\cdot\rangle_{_\mathcal{H}}$ is called a reproducing kernel Hilbert space (RKHS) if there exists a bivariate function $K(\cdot\,,\,\cdot)$ on $\mathcal{T}\times \mathcal{T}$, called a reproducing kernel, such that for every $t\in \mathcal{T}$ and $f\in \mathcal{H}$,
\begin{itemize}
\item[(i) ] $K(\cdot,\,t)\in \mathcal{H}$,
\item[(ii) ] $f(t)=\langle\,f,\,K(\cdot,\,t)\,\rangle_{_\mathcal{H}}$. 
\end{itemize}
Relation (ii) is called the reproducing property of $K$. The reproducing kernel of an RKHS is nonnegative definite and unique and conversely a nonnegative definite function uniquely determines an RKHS. 

Methodologies based on RKHS have been used extensively in the literature on nonparametric regression and function estimation. For example, in the nonparametric regression framework, let $\mathcal{H}:=\mathcal{W}_2^r$, where $\mathcal{W}_2^r$ is the $r^\text{th}$ order Sobolev Hilbert space defined by
$$
\mathcal{W}_2^r = \left\{ g:[0,1] \to \mathbb{R} \,\,\Big|\,\,g,g^{(1)},...,g^{(r - 1)} \text{ are absolutely continuous and } g^{(r)} \in L^2([0,1]) \right\}.
$$
Here $g^{(s)}$ denotes the $s^\text{th}$ derivative of the function $g$. If we endow $\mathcal{W}_2^r$ with the squared norm 
$$\|g\|^2_{\mathcal{W}_2^r}=\sum \limits_{k=0}^{r-1} \left( \int_0^1 g^{(k)}(t)\,dt \right) ^2 + \int_0^1 [g^{(r)}(t)]^2\,dt\,$$ then $\mathcal{W}_2^r$ is an RKHS with the reproducing kernel
$$
K_r (s,t) = \frac{1}{(r!)^2}\,B_r(s)\,B_r(t) + \frac{( - 1)^{r - 1}}{(2r)!}\,B_{2\,r}(|s - t|)\,,
$$
where $B_r(\cdot)$ is the $r^\text{th}$ Bernoulli polynomial. Therefore the regularized estimator of the unknown regression function on $\mathcal{W}_2^r$ with the penalty functional $\int_0^1 [g^{(r)}(t)]^2\,dt$ coincides with the usual smoothing spline estimator. See Wahba (1990).

From now on, we assume that the sample paths of the processes $X$ and $Y$ belong to an RKHS $\mathcal{H}$ almost surely. 
In addition, we assume that the reproducing kernel $K$ is square integrable. Therefore, by Mercer's theorem
\begin{equation}
K(s,\,t) = \sum\limits_{k \geq 1} \rho_{_k}\, \varphi_{_k} (s)\,\varphi_{_k} (t) \,.
\end{equation}
where $\rho_{_1},\,\rho_{_2},\,\dots$ are constants and $\varphi_{_1},\,\varphi_{_2},\,\dots$ are orthonormal basis for $L_2(\mathcal{T})$, that is
\begin{equation*}
\langle\, \varphi_{_j} ,\,\varphi_{_k}\,\rangle_{_{L^2(\mathcal{T})}} = \int_{\mathcal{T}} \varphi_{_j} (t)\,\varphi_{_k} (t)\,dt = \delta _{j\,k}\,,
\end{equation*}
and $\delta $ is Kronecker's delta. So we have the following representations for $X$ and $Y$
\begin{equation}\label{RXY}
X(\cdot) = \sum\limits_{k \geq 1} x_{_k} \,\varphi_{_k} (\cdot),\,\qquad 
Y(\cdot) = \sum\limits_{k \geq 1} y_{_k}\, \varphi_{_k} (\cdot)\,,
\end{equation}
where $x_{_k}$ and $y_{_k}$ are the respective random coefficients. 
Now from Lemma 1.1.1 in Wahba (1990), we know that any square integrable function $f$ on $\mathcal{T}$ belongs to $\mathcal{H}(K)$ if and only if
\begin{equation} \label{Wahba}
\|\,f\,\|^2_{_{\mathcal{H}(K)}}=\sum\limits_{k \geq 1}\rho_{_k}^{-1}\,f^2_{_k}<\infty\,,
\end{equation}
where for $k=1,\,2,\,\dots$ we have $f_{_k}=\langle\,f,\,\varphi_{_k}\,\rangle_{_{L_2(\mathcal{T})}}$.

For Hilbert spaces $\mathcal{H}_1$ and $\mathcal{H}_2$ with corresponding inner products $\langle\cdot,\,\cdot\rangle_{_{\mathcal{H}_1}}$ and $\langle \cdot,\,\cdot\rangle_{_{\mathcal{H}_2}}$, the tensor product Hilbert space of $\mathcal{H}_1$ and $\mathcal{H}_2$ is defined in the following fashion. Let $f_1,\,f_2\in \mathcal{H}_1$ and $g_1,\,g_2\in \mathcal{H}_2$, define $(f_1\otimes g_1)(s,\,t)=f_1(s)\,g_1(t)$ and $\mathcal{H}:=\mathcal{H}_1 \otimes \mathcal{H}_2= \lbrace h\,;\,\,h=f \otimes g,\,\,\text{for } f\in\mathcal{H}_1 , g\in \mathcal{H}_2 \rbrace$. It is easy to show that $\mathcal{H}$ is a Hilbert space with the following inner product
\begin{equation*}
\langle \,f_1\otimes g_1,\,f_2\otimes g_2\,\rangle_{_\mathcal{H}}=\langle \,f_1,\,f_2\,\rangle_{_{\mathcal{H}_1}}\langle \,g_1,\,g_2\,\rangle_{_{\mathcal{H}_2}}\,,
\end{equation*}
and $\mathcal{H}$ is called the tensor product Hilbert space of $\mathcal{H}_1$ and $\mathcal{H}_2$.  

Now consider the tensor product Hilbert space $\mathcal{H}(K\otimes K):=\mathcal{H}(K)\otimes \mathcal{H}(K)$. For $f,\,g\in \mathcal{H}(K)$ and $s,\,t\in \mathcal{T}$ we have
\begin{align*}\label{l}
(f \otimes g)(s,\,t) &= f(s)\,g(t)\\
&=  \langle\, f,\,K(\cdot,s) \rangle_{_{\mathcal{H}(K)}}\,\langle\, g,\,K(\cdot,\,t)\,\rangle_{_{\mathcal{H}(K)}}\\
&= \langle\, f \otimes g,\,K(\cdot,\,s) \otimes K(\cdot,\,t)\,\rangle_{_{\mathcal{H}(K \otimes K)}}\,.
\end{align*}
This implies that $\mathcal{H}(K\otimes K)$ is an RKHS with the reproducing kernel
\begin{equation*}
(K\otimes K)((s_1,\,t_1),\,(s_2,\,t_2))=K(s_1,\,s_2)\,K(t_1,\,t_2)\,.
\end{equation*}
A result in Cai and Yuan (2010) states that if $E\left[\|X\|^2_{_{\mathcal{H}(K)}}\right]<\infty$ then $\mu _{_X}\in \mathcal{H}(K)$ and $C_{_{XX}}\in \mathcal{H}(K\otimes K)$. In addition, if  we assume that  $E\left[\|Y\|^2_{_{\mathcal{H}(K)}}\right]<\infty $ then we have the following result. 

\begin{theorem}\label{T1}
If $E\left[ \|X \otimes Y\|^2_{_{\mathcal{H}(K \otimes K)}}\right]<\infty $ then $C_{_{X\,Y}}$, $C_{_{Y\,X}}$, $A_{_{X\,Y}}$ and $A_{_{Y\,X}}$ all belong to the tensor product Hilbert space $\mathcal{H}(K\otimes K)$.
\end{theorem}
Theorem \ref{T1} is fundamental and will be used in Section \ref{EP} to derive a regularized estimator of the cross-covariance function. 
\section{Estimation procedures}\label{EP}
Recall the functional linear regression \eqref{flr1.1}, that is
\begin{equation*}
Y(t) =\alpha (t)+ \int_\mathcal{T} X(s)\,\beta (s,\,t)\,ds  + \eta (t)\,,
\end{equation*}
where $\beta $ is an unknown square integrable slope coefficient function, $\alpha$ is an intercept function and $\eta$ is a noise term with zero mean and finite variance.
Without loss of generality, we assume that both processes $X$ and $Y$ are centred, that is $E\left[X(s)\right]=0$ and $E\left[Y(t)\right]=0$. Therefore, the functional linear model \eqref{flr1.1} can be rewritten as
\begin{equation*}
Y(t) = \int_\mathcal{T} X(s)\,\beta (s,\,t)\,ds  + \eta (t)\,.
\end{equation*}
He \textit{et al.} (2000) showed that under certain regularity conditions, the bivariate coefficient function $\beta$ admits the following representation
\begin{equation}\label{flr3}
\beta(s,t) = \sum\limits_{k = 1}^\infty  \sum\limits_{\ell = 1}^\infty  \frac{\sigma_{_{k\,\ell}}}
{\lambda _{_{X\,\ell}}} \,\Psi _{_\ell} (s)\,\Phi _{_k} (t)\,,
\end{equation}
where $\sigma _{_{k\,\ell}}=E\left[\zeta _{_\ell} \xi _{_k} \right]$ and $\zeta_{_\ell}$, $\xi _{_k} $ are given in \eqref{KLexp}. In addition, if we assume that 
\begin{equation}\label{siglam}
\sum\limits_{k = 1}^\infty  \sum\limits_{\ell = 1}^\infty  \frac{\sigma^2_{_{k\,\ell}}}{\lambda^2_{_{X\,\ell}}}<\infty\,,
\end{equation}
then Lemma A.2 in Yao {\it et al.} (2005b) implies that the right hand side of \eqref{flr3} converges in the $L_2$-sense.

In this section, we utilize a regularization approach to estimate the cross-covariance function of $X$ and $Y$ by assuming that the sample paths of $X$ and $Y$ are smooth in the sense that they belong to a certain RKHS. As a by-product of this, we provide estimates of the singular value functions and also the coefficient function $\beta$ using the representation \eqref {flr3}. 

\subsection{Cross covariance and singular components functions}
In this section, we employ a regularization method similar to that in Cai and Yuan (2010) to estimate the cross-covariance function and then use it to propose estimates of the regression coefficient function $\beta$ and functional singular components. First note that from Theorem \ref{T1} we have $C_{_{X\,Y}}\in \mathcal{H}(K\otimes K)$. We propose estimating $C_{_{X\,Y}}$ by a function $\widehat{C}_{_{X\,Y}}$ that minimizes
\begin{equation}\label{E(C)}
E(C)  = \ell _n (C) + \lambda\, \|C\|_{_{\mathcal{H}(K \otimes K)}}^2\,, \qquad \lambda \geq 0\,,
\end{equation}
over the space $\mathcal{H}(K\otimes K)$.  
Here the objective function $\ell _n (C)$ is defined as an average of $n$ scaled squared Frobenius norms of differences between two $m_{_1}\times m_{_2}$ matrices.  For each $i=1,\,\dots,\,n$, one matrix has the entries given by $\left[U_{i\,j_{_1}}-\mu_{_X}(S_{i\,j_{_1}})\right]\,\left[V_{i\,j_{_1}}-\mu_{_Y}(T_{i\,j_{_2}})\right]$ and the other one with entries $C(S_{i\,j_{_1} } ,T_{i\,j_{_2} } )$, where $j_{_1}=1,\,\dots,\,m_{_1}$ and $j_{_2}=1,\,\dots,\,m_{_2}$. More formally    
\begin{equation*}
\ell _n (C) = \frac{1}
{n\,m_{_1}\, m_{_2} }\sum\limits_{i = 1}^n \sum\limits_{j_{_1}  = 1}^{m_{_1} } \sum\limits_{j_{_2}  = 1}^{m_{_2} } \left\lbrace \left[U_{i\,j_{_1}}-\mu_{_X}(S_{i\,j_{_1}})\right]\,\left[V_{i\,j_{_1}}-\mu_{_Y}(T_{i\,j_{_2}})\right] - C(S_{i\,j_{_1} } ,T_{i\,j_{_2} } )\right\rbrace^2  \,.
\end{equation*}
Note that the expression \eqref{E(C)} represents the tradeoff between the goodness of fit measured by $\ell _n$ and smoothness of the solution measured by the RKHS norm $\|C\|_{_{\mathcal{H}(K \otimes K)}}^2$. 
The mean functions $\mu_{_X}$ and $\mu_{_Y}$ are often unknown in practice, so it is necessary to replace $\mu _{_X}(S_{i\,j_{_1}})$ and $\mu _{_Y}(T_{i\,j_{_1}})$ in $\ell _n (C)$ by their estimates, denoted by $\widehat{\mu}_{_X} (S_{i\,j_{_1} })$ and $\widehat{\mu}_{_Y} (T_{i\,j_{_2} })$, whenever needed. Therefore a more realistic definition of $\ell _n (C)$ is 
\begin{equation}\label{ln}
\ell _n (C) = \frac{1}
{n\,m_{_1}\, m_{_2} }\sum\limits_{i = 1}^n \sum\limits_{j_{_1}  = 1}^{m_{_1} } \sum\limits_{j_{_2}  = 1}^{m_{_2} } \left\lbrace C_i(S_{i\,j_{_1}},\,T_{i\,j_{_2}}) - C(S_{i\,j_{_1} } ,\,T_{i\,j_{_2} } )\right\rbrace^2 \,,
\end{equation}
where
\begin{equation*}
C_i(S_{i\,j_{_1}},T_{i\,j_{_2}})=\left[U_{i\,j_{_1} }  - \widehat{\mu}_{_X} (S_{i\,j_{_1} } )\right]\,\left[V_{i\,j_{_2} }  - \widehat{\mu}_{_Y} (T_{i\,j_{_2} } )\right]\,.
\end{equation*}
Under the random design setup discussed in this paper, the mean function estimates $\widehat{\mu}_{_X} (S_{i\,j_{_1} })$ and $\widehat{\mu}_{_Y} (T_{i\,j_{_2} })$ can be obtained using any of the methods discussed in Yao {\it et al.} (2005a), Li and Hsing (2010), and Cai and Yuan (2011). Because the method of Cai and Yuan (2011) has an optimal rate of convergence under $L^2$ norm and the current paper pursues optimality and the rate of convergences in the $L^2$ sense, it is natural for us to employ the method that was discussed in Cai and Yuan (2011) for estimating the mean functions.
 
The solution of the optimization problem in \eqref{E(C)} can be obtained by the following version of the so called the representer lemma. See for example Wahba (1990). 
\begin{lemma}\label{L1}
The solution $\widehat{C}_{_{X\,Y}}$ of the minimization problem in \eqref{E(C)} has the following form
\begin{equation}\label{represent}
\widehat{C}_{_{X\,Y}}(s,t) = \sum\limits_{i = 1}^n \sum\limits_{j_{_1} = 1}^{m_{_1}} \sum\limits_{j_{_2} = 1}^{m_{_2}} a_{i\,j_{_1}\, j_{_2} } K(s,\,S_{i\,j_{_1}} )\,K(t,\,T_{i\,j_{_2}}) 
\end{equation}
where $a_{i\,j_1,\,j_2}$ is the solution of the equation
\begin{equation*}
\sum\limits_{i = 1}^n \sum\limits_{j_{_1}  = 1}^{m_{_1} } \sum\limits_{j_{_2}  = 1}^{m_{_2} } \left\lbrace K(S_{r\,k} ,\,S_{i\,j_{_1} } )\,K(T_{r\,\ell} ,T_{i\,j_{_2} } ) + \lambda\, n\,m_{_1}\, m_{_2}\, \delta _{i\,r} \,\delta _{j_{_1} k} \,\delta_{j_{_2} \ell}\right\rbrace a_{i\,j_{_1} \,j_{_2} }= C_r (S_{r\,k} ,T_{r\,\ell} )
\end{equation*}
or equivalently
\begin{equation}\label{subject}
\frac{1}
{n\,m_1 \,m_2 }\left\lbrace C_r (S_{r\,k} ,T_{r\,\ell}) - \widehat{C}_{_{X\,Y}} (S_{r\,k} ,\,T_{r\,\ell} )\right\rbrace = \lambda\, a_{r\,k\,\ell}  \,.
\end{equation}
\end{lemma}
\begin{remark}
Note that one may use penalties other than $\|\cdot \|^2_{\mathcal{H}(K\otimes K)}$ in equation \eqref{E(C)}. For example, let $J(C)$ be a penalty functional such that its null space 
$$
\mathcal{N}_0=\{C\in \mathcal{H}(K\otimes K) \, | \, J(C)=0\}
$$
is a finite dimensional subspace. Let $n_0$ denote the dimensionality of $\mathcal{N}_0$, i.e. $n_0=\operatorname{dim}(\mathcal{N}_0)$, and $\{\vartheta_1,\dots ,\vartheta_{n_0}\}$ be the orthonormal basis of $\mathcal{N}_0$. It can be shown that there exist constants $d_{\ell}$, $\ell =1,\dots ,n_0$, and $a_{ij_1j_2}$, for $j_1 =1,\dots ,m_1$, $j_2 =1,\dots ,m_2$, $i =1,\dots ,n$ such that
$$
\widehat{C}_{XY}(s,t)= \sum \limits_{\ell =1}^{n_0}d_{\ell}\vartheta_{\ell}(s,t) + \sum \limits_{i=1}^n \sum \limits_{j_1=1}^{m_1} \sum \limits_{j_2=1}^{m_2} a_{ij_1j_2}K(s,S_{ij_1})K(t,T_{ij_2})\,.
$$
As a special case, if we set $\mathcal{T}=[0,1]$, $\mathcal{H}=\mathcal{W}_2^2$ and $J(C)$ is considered as the thin plate spline penalty, that is
\[
J(C)=\int_0^1\int_0^1 \left[ \left( \dfrac{\partial ^2C}{\partial s^2} \right)^2 + 2 \left( \dfrac{\partial ^2C}{\partial s \partial t} \right)^2 + \left( \dfrac{\partial ^2C}{\partial t^2} \right)^2 \right] dsdt
\]
then we have $\mathcal{N}_0=\operatorname{span}\{1,s,t\}$.
\end{remark}
It is obvious from \eqref{represent} that this estimate satisfies $\widehat{C}_{_{Y\,X}}(t,s)=\widehat{C}_{_{X\,Y}}(s,t)$.
Now by the substitution principal, estimates of $A_{_{X\,Y}}$ and $A_{_{Y\,X}}$ are given by
\begin{equation*}
\widehat{A}_{_{X\,Y}} (s,t) = \int_\mathcal{T} {\widehat{C}_{_{X\,Y}} (s,u) \,\widehat{C}_{_{X\,Y}} (t,\,u)\,du},\quad \text{ and } \quad
\widehat{A}_{_{Y\,X}} (s,t) = \int_\mathcal{T} {\widehat{C}_{_{X\,Y}} (u,s)\, \widehat{C}_{_{X\,Y}} (u,t)\,du}\,.
\end{equation*}

\begin{remark}
From Theorem \ref{T1}, we know that $A_{_{X\,Y}}$ and $A_{_{Y\,X}}$ belong to the tensor product space $\mathcal{H}(K\otimes K)$. This suggests that we can use a similar regularization method in \eqref{E(C)} to estimate $A_{_{X\,Y}}$ and $A_{_{Y\,X}}$, as an alternative to the aforementioned subtitution method.
\end{remark}
Since $(\sigma _k^2,\psi _k)$ and $(\sigma _k^2,\phi _k)$ are eigenvalues-eigenfunctions of the operators $\mathcal{A}_{X\,Y}$ and $\mathcal{A}_{Y\,X}$, respectively, the representation \eqref{represent} suggests a simple method to estimate $(\sigma_k^2,\,\psi_k,\,\phi_k)$. In what follows we discuss this method. We use a setup similar to that in Cai and Yuan (2010). Consider the $n\,m_{_1} \times n\,m_{_2}$ block diagonal matrix 
\begin{equation*}
\mathbf{A} = \begin{pmatrix}
 \mathbf{A}_1 & \mathbf{0}& \mathbf{0}& \cdots & \mathbf{0}\\
\mathbf{0} & \mathbf{A}_2 & \mathbf{0} & \cdots &\mathbf{0} \\
 \vdots &\vdots &\ddots &\vdots &\vdots \\
\mathbf{0} &\mathbf{0}& \mathbf{0}&\ldots & \mathbf{A}_n  
\end{pmatrix} \,,
\end{equation*}
where for $i=1,\,\dots,\,n$, the $m_{_1}\times m_{_2}$ matrices $\mathbf{A}_{_i}$ are given by $\mathbf{A}_i=\left(\,a_{i\,j_{_1}\,j_{_2}}\,\right)_{1\leq j_1\leq m_{_1},\,1\leq j_{_2}\leq m_{_2}}$. Also define the $m_{_1}\,n\times m_{_1}\,n$ matrix $\mathbf{P}$ in the following form 
\begin{equation*}
\mathbf{P} = \begin{pmatrix}
 \mathbf{P}_{11} &  \mathbf{P}_{12}&  \mathbf{P}_{13} &\cdots &  \mathbf{P}_{1n}  \\
 \mathbf{P}_{21} &  \mathbf{P}_{22} &  \mathbf{P}_{23} &\cdots&  \mathbf{P}_{2n} \\
 \vdots &\vdots &\ddots &\vdots &\vdots \\
 \mathbf{P}_{n1} & \mathbf{P}_{n2} & \mathbf{P}_{n3}&\ldots & \mathbf{P}_{nn}  \\ 
\end{pmatrix}
\end{equation*}
where
\begin{equation*}
 \mathbf{P}_{i_{_1} i_{_2} }  = \left( \int_\mathcal{T} K(s,\,S_{i_{_1}\, j_{_1} } )\, K(s,\,S_{i_{_2} j_{_2}} )\,ds \right)_{1 \leq j_{_1} ,\,j_{_2}  \leq m_{_1}},\,\qquad \text{ for } \quad  1 \leq i_{_1} ,\,i_{_2}  \leq n\,.
\end{equation*}
Similarly define the $m_{_2}\,n\times m_{_2}\,n$ matrix
\begin{equation*}
\mathbf{Q} = \begin{pmatrix}
 \mathbf{Q}_{11} &  \mathbf{Q}_{12}&  \mathbf{Q}_{13} &\cdots &  \mathbf{Q}_{1n}  \\
 \mathbf{Q}_{21} &  \mathbf{Q}_{22} &  \mathbf{Q}_{23} &\cdots&  \mathbf{Q}_{2n} \\
 \vdots &\vdots &\ddots &\vdots &\vdots \\
 \mathbf{Q}_{n1} & \mathbf{Q}_{n2} & \mathbf{Q}_{n3}&\ldots & \mathbf{Q}_{nn}  \\ 
\end{pmatrix}
\end{equation*}
where
\begin{equation*}
\mathbf{Q}_{i_1 i_2 }  = \left( \int_\mathcal{T} K(s,\,T_{i_{_1}\, j_{_1} } )\, K(s,\,T_{i_{_2} j_{_2}} )\,ds \right)_{1 \leq j_{_1} ,\,j_{_2}  \leq m_{_2}},\,\qquad \text{ for } \quad  1 \leq i_{_1} ,\,i_{_2}  \leq n\,.
\end{equation*}
Using these notations, we have the following lemma.
\begin{lemma}
The singular functions $\psi_{_k}$ and $\phi_{_k}$ can be estimated by
\begin{equation*}
\widehat{\psi}_{_k} (\cdot) = \bfmath{\alpha}_{_k} ^\prime\,  \mathbf{g}_{_1} (\cdot)\,,
\end{equation*}
\begin{equation*}
\widehat{\phi}_{_k} (\cdot) = \bfmath{\beta}_{_k}^\prime\,  \mathbf{g}_{_2} (\cdot)\,,
\end{equation*}
where $\bfmath{\alpha}_{_k}$ and $\bfmath{\beta}_{_k}$ are the k-th columns of $\mathbf{P}^{- 1/2}\, \mathbf{W}_{_1}$ and $\mathbf{Q}^{-1/2}\,\mathbf{W}_{_2}$, respectively. Here 
$\mathbf{W}_{_1}$ and $\mathbf{W}_{_2}$ are two matrices with their columns being the eigenvectors of 
 $\mathbf{P}^{1/2}\,\mathbf{A}\,\mathbf{Q}\,\mathbf{A}'\,\mathbf{P}^{1/2}$
and $\mathbf{Q}^{1/2}\,\mathbf{A}'\,\mathbf{P}\,\mathbf{A}\,\mathbf{Q}^{1/2}$, respectively, and also 
\begin{align*}
\mathbf{g}_{_1}(\cdot)&=(\,K(\cdot,\,S_{1\,1}),\,\dots,\,K(\cdot,\,S_{1\,m_1}),\,K(\cdot,\,S_{2\,1}),\,\dots,\,K(\cdot,\,S_{2\,m_1})\,,\dots,\,K(\cdot,\,S_{n\,1}),\,\dots,\,K(\cdot,\,S_{n\,m_1})\,)'\,,\\
\mathbf{g}_2(\cdot)&=(\,K(\cdot,\,T_{1\,1}),\,\dots,\,K(\cdot,\,T_{1\,m_2}),\,K(\cdot,\,T_{2\,1}),\,\dots,\,K(\cdot,\,T_{2\,m_2}),\,\dots,\, K(\cdot,\,T_{n\,1}),\,\dots,\,K(\cdot,T_{n\,m_2}))'.
\end{align*}
\end{lemma}
\subsection{Estimating the regression function}
From the representation \eqref{flr3}, in order to estimate the coefficient function $\beta$ we require to estimate $\lambda _{_{X\,\ell}}$, $\sigma _{_{k\,\ell}}$, $\Psi_{_\ell}$ and $\Phi_{_k}$. Cai and Yuan (2010) suggested estimates of $\lambda _{_{X\,\ell}}$, $\Psi _\ell$ and $\Phi _k$. So it remains to estimate $\sigma _{_{k\,\ell}}$. Notice that the expansion 
\begin{equation*}\label{sigmakl}
C_{_{X\,Y}}(s,\,t) = \sum\limits_{k = 1}^\infty \sum\limits_{\ell = 1}^\infty  \sigma_{_{k\,\ell}}\,\Psi_\ell(s)\,\Phi _k(t)
\end{equation*}
implies the representation 
\begin{equation*}
\sigma_{_{k\,\ell}} = \int_{\mathcal{T} \times \mathcal{T}}\Psi_{_\ell}(s)\,C_{_{X\,Y}}(s,t)\,\Phi_k(t)\,ds\,dt\,. 
\end{equation*}
which in turn, using the estimates $\widehat{\Psi}_{_\ell}, \,\widehat{\Phi}_{_k}$ in Cai and Yuan (2010) and $\widehat{C}_{_{X\,Y}}$ in Lemma \ref{L1}, we obtain 
\begin{equation*}
\widehat{\sigma}_{_{k\,\ell}} = \int_{\mathcal{T} \times \mathcal{T}} \widehat{\Psi}_{_\ell}(s)\,\widehat{C}_{_{X\,Y}}(s,t)\,\widehat{\Phi}_{_k}(t)\,ds\,dt\,. 
\end{equation*}
Therefore an estimate of the coefficient function $\beta$ is given by
\begin{equation}\label{slophat}
\widehat{\beta}(s,\,t) = \sum\limits_{k = 1}^{J_{_1}} \sum\limits_{\ell = 1}^{J_{_2}} \frac{\widehat{\sigma}_{_{k\,\ell}}}
{\widehat{\lambda}_{_{X\,\ell}}}\,\widehat{\Psi}_\ell(s)\,\widehat{\Phi}_k(t)\,, 
\end{equation}
where the numbers $J_{_1}$ and $J_{_2}$ must be determined. One may use the methods such as AIC and BIC to choose $J_{_1}$ and $J_{_2}$. See for example Yao {\it et al.} (2005a \& 2005b).

\section{Rates of convergence}\label{S4}
In this section we derive the convergence rates of the cross covariance function estimator $\widehat{C}_{_{X\,Y}}$ and singular components. We also obtain the theoretical properties and convergence rate of $\widehat{\beta}$. We consider assumptions similar to those in Cai and Yuan (2010). Let $\mathcal{F}(\nu ;\,M,\,c)$ be the collection of probability measures defined on $(X,Y)$ such that
\begin{itemize}
\item[(a)] the sample paths of $\nu$-time differentiable processes $X$ and $Y$ belong to $\mathcal{H}(K)$ almost surely and 
 $E\left[\|X\otimes Y\|^2_{_{\mathcal{H}(K\otimes K)}}\right]<M$,
\item[(b)] $K$ is a Mercer kernel with eigenvalues $\rho_{_k}$, satisfying $\rho_{_k}\asymp k^{-2\,\nu}$, where for two positive sequences $r_n$ and $s_n$, the notation $r_n\asymp s_n$ means that
 $$0 < \mathop{\lim \inf }\limits_{n \to \infty } (r_n/s_n) \le \mathop{\lim \sup }\limits_{n \to \infty } (r_n/s_n) < \infty \,.$$
\item[(c)] For the process $Z$ being either $X$ or $Y$, there exists a constant $c>0$ such that 
$$E\left[Z^4(t)\right]\leq c\,\left(E\left[Z^2(t)\right]\right)^2\quad \text{ for any }\quad t\in \mathcal{T},\,$$
and
\begin{equation*}
E\left[ \int_\mathcal{T}Z(t)\,f(t)\,dt\right]^4 \le {c}\,\left(E\left[\int_\mathcal{T}Z(t)\,f(t)\,dt\right]^2 \right)^2,
\end{equation*}
for any $f\in L_2(\mathcal{T})$.
\end{itemize}
The condition (a) imposes smoothness of the processes $X$ and $Y$ and therefore $X\otimes Y$. The boundedness requirement $E\left[\|X\otimes Y\|^2_{_{\mathcal{H}(K\otimes K)}}\right]<M$ in (a) is a technical condition.  The condition (b) guaranties the smoothness of the kernel function $K$. Finally, the condition (c) is also technical which concerns the fourth moments of both processes $X$ and $Y$. Note that (c) is satisfied with $c=3$ when processes $X$ and $Y$ are both Gaussian.

Let $m$ be the harmonic mean of $m_{_1}$ and $m_{_2}$, i.e.,
$$
m = \left( \frac{1}
{2\,m_{_1}} + \frac{1}
{2\,m_{_2}} \right)^{ - 1} \,.
$$
For the tuning parameter $\lambda$, we assume that
\begin{equation}\label{lambda}
\lambda \asymp 
\left(\frac{\log(n)}
{m\,n} \right)^{\frac{2\,\nu}
{2\,\nu  + 1}}\,
\end{equation}

\noindent
Corollary 5 in Cai and Yuan (2010) states that for any $\nu$ times differentiable process $Z$ that almost surely belongs to an RKHS and satisfies the condition
\begin{equation}\label{meanC}
\|\widehat \mu_{_Z}  - \mu_{_Z}\|_{_{L_2 }}^2  = O_p \left( \left(\frac{\log(n)}
{m_z\,n} \right)^{\frac{2\,\nu}
{2\,\nu  + 1}}  + \frac{1}
{n} \right),
\end{equation}
we have 
\begin{equation}\label{PsiCR}
\|\widehat \Psi_{_k} - \Psi_{_k}\|_{_{L_2}}^2  = O_p \left(\left(\frac{\log(n)}
{m_z\,n} \right)^{\frac{2\,\nu }
{2\,\nu  + 1}} + \frac{1}{n} \right),\quad \text{for } k\leq J \text{ with } J \text{ fixed}.
\end{equation}
Notice that in \eqref{meanC} and \eqref{PsiCR}, we set $m_{_z}=m_1$ if $Z$ is replaced by $X$ and $m_{_z}=m_2$ if $Z$ is replaced by $Y$. 
The following result gives the rate of convergence for $\widehat{C}_{_{X\,Y}}$ in terms of the integrated squared error loss.

\begin{theorem}\label{T3}
Assume that $E\left[\varepsilon^4 \right]<\infty$ and $E\left[\epsilon ^4\right]<\infty$, and random times points $S_{_{i\,j}}$ and $T_{_{i\,j}}$ are independent and identically distributed with a common density function bounded away from zero on $\mathcal{T}$. In addition, assume that the estimators $\widehat{\mu}_{_X}(\cdot)$ and $\widehat{\mu}_{_Y}(\cdot)$ satisfy \eqref{meanC} and the tuning parameter $\lambda$ satisfies \eqref{lambda}. Then
\begin{equation}
\mathop {\lim }\limits_{D \to \infty } \mathop {\lim \sup }\limits_{n \to \infty } \mathop {\sup }\limits_{F \in \mathcal{F}(\nu;\,M ,\,c )} P_{_F}\left\lbrace \left\|\widehat{C}_{_{X\,Y}}  - C_{_{X\,Y}} \right\|_{_{L_2 }}^2  > D\left[\left(\frac{\log( n)}
{m\,n} \right)^{\frac{2\,\nu }
{2\,\nu  + 1}}  + \frac{1}
{n}\right] \right\rbrace= 0.
\end{equation}
\end{theorem}
The proof of Theorem \ref{T3} is similar to that of Theorem 4 in Cai and Yuan (2010) and will be omitted.

\begin{corollary}\label{Cor1}
Under the conditions of Theorem \ref{T3},
\begin{equation}\label{AXYRC}
\mathop {\lim }\limits_{D \to \infty } \mathop {\lim \sup }\limits_{n \to \infty } \mathop {\sup }\limits_{F \in \mathcal{F}(\nu ;\,M ,\,c )} P_{_F}\left\lbrace \left\|\widehat{A}_{_{X\,Y}}  - A_{_{X\,Y}} \right\|_{_{L_2}}^2  > D\left[\left(\frac{\log(n)}
{m\,n}\right)^{\frac{2\,\nu}
{2\,\nu  + 1}}  + \frac{1}
{n} \right] \right\rbrace = 0\,,
\end{equation}
\begin{equation}\label{AYXRC}
\mathop {\lim }\limits_{D \to \infty } \mathop {\lim \sup }\limits_{n \to \infty } \mathop {\sup }\limits_{F \in \mathcal{F}(\nu ;\,M ,\,c)} P_{_F}\left\lbrace \left\|\widehat{A}_{_{Y\,X}}  - A_{_{Y\,X}} \right\|_{_{L_2 }}^2  > D\left[\left(\frac{\log(n)}{mn} \right)^{\frac{2\,\nu}
{2\,\nu  + 1}}+ \frac{1}
{n}\right] \right\rbrace = 0\,.
\end{equation}
\end{corollary}
\begin{corollary}
Under the conditions of Theorem \ref{T3}, for all $k \leq J$, where $J$ is fixed, if $\sigma _k^2$ is of multiplicity one, then
\begin{equation}\label{52}
\left|\widehat{\sigma}_k ^2 - \sigma_k ^2\right|^2 = O_p \left(\left(\frac{\log(n)}
{m\,n}\right)^{\frac{2\,\nu}
{2\,\nu  + 1}}  + \frac{1}
{n} \right)\,,
\end{equation}
\begin{equation}\label{50}
\left\|\widehat{\phi}_k  - \phi_k \right\|_{_{L_2}}^2 = O_p \left(\left(\frac{\log(n)}
{m\,n}\right)^{\frac{2\,\nu}
{2\,\nu  + 1}}  + \frac{1}
{n} \right)\,,
\end{equation}
\begin{equation}\label{51}
\left\|\widehat \psi_k  - \psi_k \right\|_{_{L_2}}^2 = O_p \left(\left(\frac{\log(n)}
{m\,n}\right)^{\frac{2\,\nu}
{2\,\nu  + 1}}  + \frac{1}
{n} \right)\,.
\end{equation}
\end{corollary}

\noindent
Now, we are in a position to obtain the convergence rate of  $\widehat{\beta}$. First we assume that for some $\tau , \gamma >0$,
\begin{equation}\label{sig}
\left(\frac{\sigma_{k\,\ell}}
{\lambda_{_{X\,\ell}}} \right)^2  \leq c_{_0}\, k^{ -(\tau +1)} \ell^{ - (\gamma  + 1)}\,, 
\end{equation}
and
\begin{equation}\label{K,M}
J_{_1},\,J_{_2} \asymp\left[ \left(\frac{\log(n)}{m\,n} \right)^{\frac{2\,\nu}{2\,\nu  + 1}}+\frac{1}{n} \right]^{ - \frac{1}{\tau  + \gamma  + 2}}\,.
\end{equation}
\begin{remark}
It may seem that $\tau$ and $\gamma$ should be related to $\nu$. So it is worthwhile to discuss this a bit more. In general, $\tau$ and $\gamma$ are not functions of $\nu$. For example suppose $X(t)$ is a $\nu$-time differentiable process and $Y(t)=X(t)+L$, for a constant $L \in \mathbb{R}$. It is clear that $C_{_{XX}}(s,t)=C_{_{YY}}(s,t)=C_{_{XY}}(s,t)$. So $X$ and $Y$ have shared eigenvalues $\lambda_{_{X\,\ell}}=\lambda_{_{Y\,\ell}}=\lambda_{_\ell}$ and eigenfunctions $\Psi_{_\ell}=\Phi_{_\ell}$. This implies that 
$$
\sigma_{_{k\,\ell}}=\int_{\mathcal{T}\times \mathcal{T}}\Psi_{_\ell}(s)C_{_{XY}}(s,t)\Phi_{_k}(t)ds\,dt=\lambda_{_k} \delta_{_{k\,\ell}},
$$
where $\delta$ is Kronecker's delta. Therefore for $k=\ell$, condition \eqref{sig} does not hold because $k^{\tau +\gamma +2}$ is not bounded for every $k$, meaning that, for a given $\nu$ one can not find any $\tau$ and $\gamma$ satisfying condition \eqref{sig}. In contrast, suppose $X(t)$ and $Y(t)$ are two uncorrelated $\nu$-time differentiable processes. In this case $C_{_{XY}}(s,t)=0$, for any $s,\,t \in \mathcal{T}$ and then $\sigma_{k\ell}=0$. Therefore, condition \eqref{sig} holds for every $\tau,\,\gamma >0$, meaning that for a fixed $\nu$, any $\tau$ and $\gamma$ satisfy the condition \eqref{sig}. 

\end{remark}
\begin{remark}
Assumption \eqref{sig} implies \eqref{siglam} which in turn says that the sequence     
$$\lbrace \sigma^2_{_{k\,\ell}}/\lambda^2_{_{X\,\ell}};\,\,k,\,\ell \geq 1\rbrace$$ converges to 0 with a polynomial rate. 
\end{remark}
The condition \eqref{sig} is essential for the subsequent results in this paper. Let $\mathcal{F}_{_0} (\nu ;\,M,\,c,\,c_{_0})$ be the set of distributions $F$ in $\mathcal{F}(\nu ;\,M,\,c)$ that satisfy \eqref{sig}. The convergence rate of $\widehat{\beta}$ is established in the following theorem.
\begin{theorem}\label{Sloprate}
Under the conditions of Theorem \ref{T3},
\begin{equation}
\mathop {\lim }\limits_{D \to \infty } \mathop {\lim \sup }\limits_{n \to \infty } \mathop {\sup }\limits_{F \in  \mathcal{F}_{_0} (\nu ;\,M,\,c,\,c_{_0})} P_F \left\lbrace\left\|\widehat{\beta}  - \beta \right\|_{_{L_2}}^2  > D\,\left[\frac{1}{n} + \left(\frac{\log(n)}{m\,n} \right)^{\frac{2\,\nu }{2\,\nu  + 1}} \right]^{\frac{\tau  + \gamma}{\tau  + \gamma  + 2}} \right\rbrace = 0\,.
\end{equation}
\end{theorem}
By Theorem \ref{Sloprate}, when the functions are densely sampled, that is $m\gg n^{\frac{1}{2\,\nu}}\log(n)$, the coefficient function can be estimated at the rate of $n^{- \frac{\tau  + \gamma }{\tau  + \gamma  + 2}}$. On the other hand, when the functions are sparsely sampled, that is $m= O\left(n^{\frac{1}{2\,\nu}}\log(n)\right)$, the rate of convergence for the estimated coefficient function changes to 
$$\left[\left(\frac{\log(n)}{m\,n} \right)^{\frac{2\,\nu }{2\,\nu  + 1}} \right]^{\frac{\tau  + \gamma}{\tau  + \gamma  + 2}}\,.$$
 
\noindent
Suppose, for some $0<\delta \leq \frac{1}{2}$,
\begin{equation}\label{BGI}
\frac{{\tau  + \gamma }}{{\tau  + \gamma  + 2}} > \frac{{(\nu  - \delta )(2\nu  + 1)}}{{\nu (2\,\nu  - 2\,\delta  + 1)}}\,.
\end{equation}
If the inequality \eqref{BGI} is satisfied and $m_1,\,m_2\asymp n^{B_\delta}$, where $B_{\delta}=\frac{(\tau +\gamma)(2\,\nu -2\,\delta +1)}{(\tau +\gamma +2)(2\,\nu -2\,\delta)}-1$, then the rate obtained in Theorem \ref{Sloprate} can not be improved.
\begin{theorem}
Suppose the inequality \eqref{BGI} holds  for some $0<\delta \leq \frac{1}{2}$, and $m_1,\,m_2\,\asymp n^{B_\delta}$, where $B_\delta=\frac{(\tau +\gamma)(2\,\nu -2\,\delta +1)}{(\tau +\gamma +2)(2\,\tau -2\,\delta)}-1$, then there exists a constant $d>0$ such that for any estimate $\widetilde{\beta} $,
\begin{equation}
\mathop{\lim \sup }\limits_{n \to \infty } \mathop{\sup }\limits_{F \in \mathcal{F}_{_0} (\nu ;\,M,\,c,\,c_{_0})} P_{_F} \left\lbrace \left\|\,\widetilde{\beta}  - \beta \,\right\|_{_{L_2}}^2  > d\,n^{ - \frac{\tau  + \gamma }{\tau  + \gamma  + 2}}\right\rbrace > 0\,.
\end{equation}
\end{theorem}
\begin{remark}\label{RM5}
In equation (1) we assumed that $m_1$ and $m_2$ are fixed and equal for all the respective curves. This assumption is not necessary and can be relaxed. Suppose the sampling frequencies are random. Let $m_{11},\dots ,m_{1n}$ and $m_{21},\dots ,m_{2n}$ are sampling frequencies of $X$ and $Y$, respectively. Suppose all the sampling frequencies are independent variables with finite variances. By the law of large numbers, all the asymptotic results holds for the harmonic means of these sampling frequencies, that is
$$
\left( \dfrac{1}{n}\sum \limits_{i=1}^n E\left[ \dfrac{1}{m_{1i}} \right]  \right)^{-1} \qquad \text{and} \qquad  \left( \dfrac{1}{n}\sum \limits_{i=1}^n E\left[ \dfrac{1}{m_{2i}} \right]  \right)^{-1}.
$$
\end{remark}
\begin{remark}\label{RM6}
One may wonder, how the results in this paper will change if the assumption that $X$ and $Y$ reside in the same space $\mathcal{H}(K)$ is violated. Let $X\in \mathcal{H}(K_{_1})$ and $Y\in \mathcal{H}(K_{_2})$ almost surely, where decaying rates $\rho_{_{1\,k}}$ and $\rho_{_{2\,k}}$ of the eigenvalues of $K_{_1}$ and $K_{_2}$ are different. Suppose $\rho_{_{1\,k}}\asymp k^{-2\nu_1}$ and $\rho_{_{2\,k}}\asymp k^{-2\nu_2}$. Therefore, it is straightforward to show that in all relevant results $\nu$ must be replaced by $\min (\nu_{_1},\nu_{_2})$. 
\end{remark}
\begin{remark}\label{RM7}
The results in the paper also holds, with some small modifications, for the case that the domain of the processes $X$ and $Y$ are not the same. In this situation, the RKHS for $X$ and $Y$ will be different and therefore Remark \ref{RM6} will be applicable. In addition, although we focused on the case that the domains of $X$ and $Y$ are compact subsets of the real line $\mathbb{R}$, the proposed methodology is also applicable to functional data on more general compact domains (for example subsets of $\mathbb{R}^d$) with minor modifications.      
\end{remark}

\section{Numerical experiments}
In this section we study the numerical performance of the proposed estimation method. We shall begin with a simulation study and then analyze a well known multicenter AIDS cohort dataset. 
\subsection{Simulation study} 
To demonstrate the performance of the estimated coefficient function $\beta$ in finite sample settings, we carried out a set of Monte Carlo experiments with different combinations of sample size and noise level. The predictor trajectories were generated independently from
$$
X(s)= \mu_X(s)+\sum \limits_{j=1}^{10}(-1)^j\,j^{-\frac{1}{2}}\,\zeta_j\,\Psi_j(s), \qquad s\in [0,1],
$$
where 
$$\Psi_j(s)=\begin{cases}
1 \qquad& \text{ for } j=1\\
\sqrt{2}\,\cos (j\,\pi s)\qquad& \text{ for }j\geq 2\,,
\end{cases}$$
the mean function $\mu_X(s)=\sum \limits_{j=1}^{10}\,4\,(-1)^j\,j^{-2}\,\Psi_j(s)$ and the random variables $\zeta_1,\,\dots ,\,\zeta_{10}$ are independently and identically distributed with the uniform distribution on $[-\sqrt{3},\sqrt{3}]$. 
Sparse and noisy observations $U_{ij}$s from random function $X$ are obtained based on equation \eqref{UV}, where $\varepsilon_{ij}$s are assumed to be independent and identically distributed $N(0,\sigma^2_{_X})$ random errors, with $\sigma^2_{_X}=({1.8031}/{\operatorname{StN}})^2$. Here $\operatorname{StN}$ refers to the signal-to-noise ratio. In addition, we assume that the random variables $S_{ij}$s are independently and identically drawn from the uniform distribution on $[0,1]$. To generate the corresponding realizations of the random response function $Y_i(t)$, we use a model of the form $${\rm E}\left[\,Y_i(t)\mid X_i\right]=\int _{0}^{1}\beta (s,\,t)\,X_i(s)\,ds\qquad \text{ for  }t\in [0,1]\,.$$ 
We consider the following two cases for the coefficient function $\beta$:
\begin{itemize}
\item[(i)]
$\beta$ is represented by the eigenfunctions of $X$ and $Y$.
\item[(ii)]
$\beta$ has a general form.
\end{itemize}
The sampling frequencies, $m_1$ and $m_2$, for both processes $X$ and $Y$ are uniformly generated from the set $\{2,3,4,5\}$. We also consider all combinations of the sample size $n\in \lbrace 25,\,50,\, 100\rbrace$ and the signal-to-noise ratio $\operatorname{StN}\in \lbrace2,\,8,\, \infty\rbrace $.

We denote our method by FRRK and compare it with the following methods: the functional linear regression via the Principal Analysis
by Conditional Estimation (PACE) algorithm in Yao {\it et al.} (2005b), and the penalized function-on-function
regression (PFFR) in Ivanescu {\it et al.} (2015). The method PFFR is implemented in the {\tt R}
package {\tt refund} (Crainiceanu {\it et al.} 2014). The method PACE is implemented in {\tt Matlab} and can be downloaded from \url{http://www.stat.ucdavis.edu/PACE/}. For both packages we use the default settings. In addition, we compute estimation errors using the mean integrated squared error, that is,
$$
\text{MISE}=\int_0^1\int_0^1 \left(\,\widehat{\beta}(s,\,t)-\beta (s,\,t)\,\right)^2\,ds\,dt,
$$
and the mean integrated absolute error, that is,
$$
\text{MIAE}=\int_0^1\int_0^1 \mid\,\widehat{\beta}(s,\,t)-\beta (s,\,t)\mid \,ds\,dt
$$
to compare the methods. All the respective two-dimensional integrals are approximated by a two-dimensional Gaussian quadrature method.

We consider two simulation cases. In Case 1, we set 
$$\Phi_j(t)=\sqrt{2}\,\sin (\,j\,\pi\, t\,)\text{ for }\,j=1,\,\dots ,\,10\,,$$ and define the coefficient function $\beta$ by
$$
\beta(s,\,t)=\sum \limits_{i=1}^{10} \sum \limits_{j=1}^{10}b_{ij}\,\Psi_i(s)\,\Phi_j(t),
$$
where $b_{ij}=3 \,i^{-\frac{1}{2}}\,2^{-i}\times(-1)^j\,j^{-2} $. For the response trajectories, the sparse and noisy observations are available through $V_{ij}={\rm E}\left[\,Y_i(T_{ij})\mid X_i\,\right] + \epsilon_{ij}$, where errors $\epsilon_{ij}$s are independent and identically distributed with $N(0,\sigma^2_{_Y})$, and $\sigma^2_{_Y}=({2.5096}/{\operatorname{StN}})^2$. In addition, the random time points $T_{ij}$s are independently and identically drawn from the uniform distribution on $[0,1]$.

Case 2 is the same as Case 1 except that we replace the coefficient function by 
$$
\beta(s,\,t)=4\,t\,e^{s-2t}\,, 
$$
and $\sigma^2_{_Y}=({1.1721}/{\operatorname{StN}})^2$.

\begin{figure}[ht!]
\centering
\includegraphics[scale=.9]{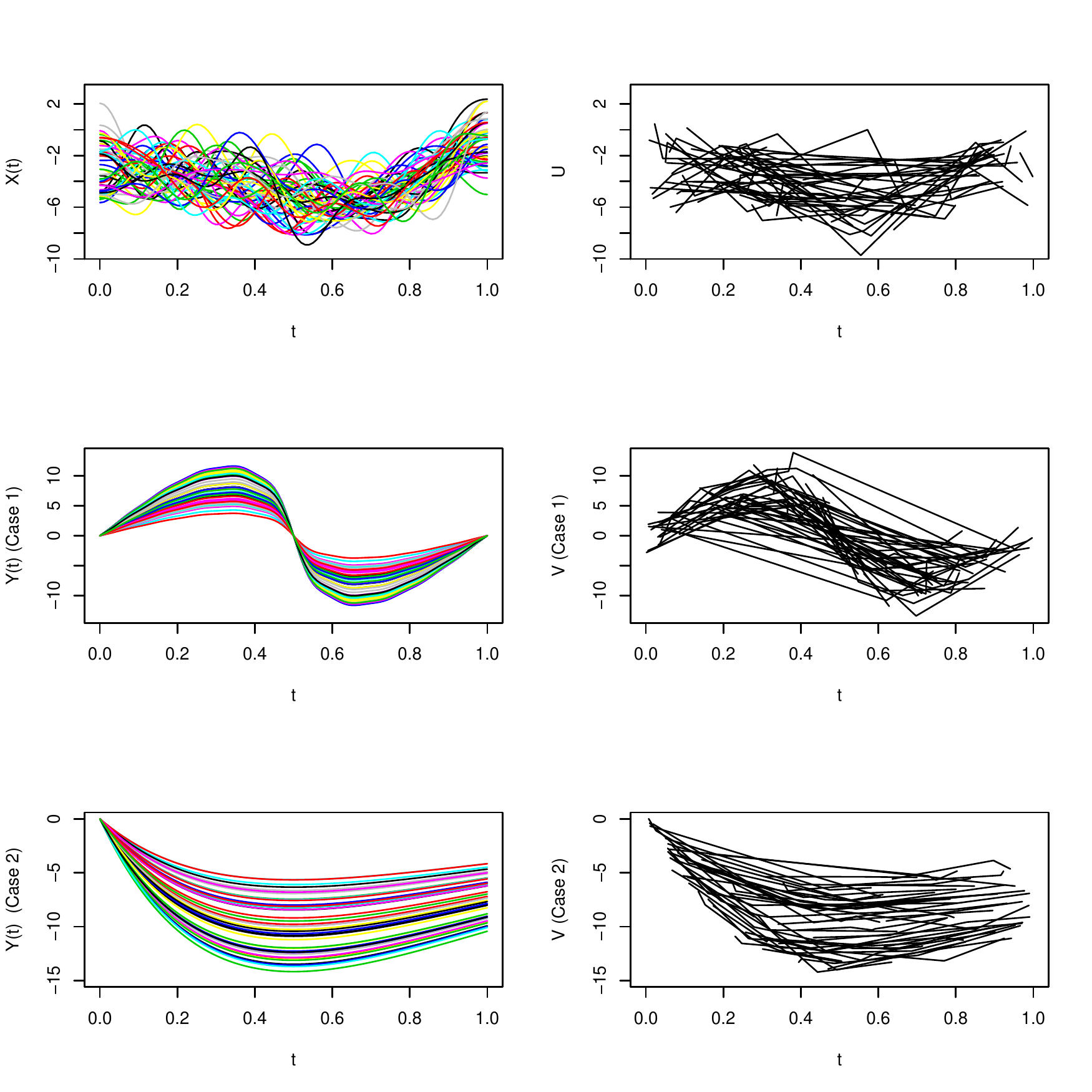}
\caption[]{The left panels give 50 simulated trajectories, the top panel for $X(s)$, the middle panel for $Y(t)$ in Case 1, and the lower panel for $Y(t)$ in Case 2. Noisy observations are shown in the right panels, the top panel for $U$, the middle panel for $V$ in Case 1, and the lower panel for $V$ in Case 2.}\label{fig1}
\end{figure}

To provide a visual inspection of typical simulated datasets from random functions $X$ and $Y$, Figure \ref{fig1} depicts fifty sampled trajectories ($n=50$). In this Figure, the top left panel is for $X$, the middle left panel for $Y$ in Case 1, and the lower left panel for $Y$ in Case 2. The right panels of Figure 1 show the corresponding observed data, that is, the top panel for $U$, the middle panel for $V$ in Case 1, and the lower panel for $V$ in Case 2.

For each combination of sample size $n$ and signal-to-noise ratio $\operatorname{StN}$, we repeat the experiment 500 times and compute MISE and MIAE of our method FRRK, the method PFFR of Ivanescu {\it et al.} (2015) and the method PACE of Yao {\it et al.} (2005b). To select the number of bases in the representation \eqref{slophat}, we use AIC for FRRK and both AIC and BIC for PACE. Table 1 and Table 2 present the Monte Carlo estimates of MISE and MIAE for these methods under Case 1 and Case 2, respectively. We see that under both cases, the proposed method FRRK (AIC) performs extremely well in compare to PFFR and PACE. In fact the estimation errors of PFFR and PACE (AIC) are dramatically large comparing to 
\begin{align*}
\text{Case 1:} &\qquad \qquad \int \beta^2(s,\,t)\,ds\,dt = 2.801542,\quad \text{ and } \quad \int \mid \beta(s,\,t)\mid\, ds\,dt= 1.417086\\ 
\text{Case 2:} &\qquad \qquad \int \beta^2(s,\,t)\,ds\,dt = 1.21695,\quad \text{ and } \quad \int \mid \beta(s,\,t)\mid\,ds\,dt = 1.020649.     
\end{align*}
This is consistent with the simulation results reported in Ivanescu {\it et al.} (2015). 

Notice that, our proposed method FRRK and the method PACE (BIC) of Yao {\it et al.} (2005b) have much smaller estimation errors in compare to PFFR and PACE (AIC) and their performance improve as the sample size $n$ or the signal-to-noise ratio $\operatorname{StN}$ increases. Among the two winners, FRRK almost uniformly outperforms PACE (BIC). It is worth mentioning that the extremely large estimation errors of PFFR and PACE (AIC) seem to appear when for some small percentages of simulated data, these methods become unstable and produce very small eigenvalue estimates. However, FRRK and PACE (BIC) appear to be more stable. In Table \ref{tab3} and Table \ref{tab4}, we remove the largest 5\% of the simulated MISE and MIAE for all methods and report the mean values of the remaining 95\% (one-sided trimmed mean). As we see, estimation errors of PFFR and PACE (AIC) decrease dramatically as expected, but still our method FRRK does a great job and uniformly outperform all the other methods by several orders of magnitude in some combinations.             
\begin{table}[h!]
\centering 
{\scriptsize \begin{tabular}{|cc|cc|cc|cc|cc|}
\hline
 & & \multicolumn{2}{c}{FRRK (AIC)} & 
\multicolumn{2}{c}{PFFR} & \multicolumn{2}{c|}{PACE (AIC)}&\multicolumn{2}{c|}{PACE (BIC)}\\
\cline{3-10}
$n$ & $\operatorname{StN}$ & MISE & MIAE & MISE & MIAE &MISE & MIAE &MISE &MIAE \\
\hline
&&&&&&&&&\\
 & $2$ & $0.464\times 10^{2}$ & $0.143\times 10^{1}$ & $0.798\times 10^{17}$ & $0.126\times 10^{8}$ & $0.424\times 10^{28}$ & $0.169\times 10^{13}$ &$0.487\times 10^{1}$ &$0.176\times 10^{1}$ \\
 &&&&&&&&&\\
\cline{3-10}
&&&&&&&&&\\
  $25$ & $8$ & $0.453\times 10^{1}$ & 0.992 & $0.296\times 10^{17}$ & $0.816\times 10^{7}$ & $0.259\times 10^6$ & $0.219 \times 10^{2}$& $0.215\times 10^{1}$&0.966 \\
  &&&&&&&&&\\
\cline{3-10}
&&&&&&&&&\\
   & $\infty$ & $0.330\times 10^{2}$ & $0.127\times 10^{1}$ & $0.158\times 10^{14}$ & $0.172\times 10^{6}$ & $0.225\times 10^{6}$ & $0.312\times 10^{2}$ &$0.192\times 10^{1}$ &0.946\\
   &&&&&&&&&\\
\hline
&&&&&&&&&\\
 & $2$ & $0.107\times 10^{1}$ & 0.755 & $0.387\times 10^{17}$ & $0.112\times 10^{8} $& $0.811\times10^{23}$ & $0.107\times 10^{11}$ &$0.130\times 10^{1}$ & 0.836\\
 &&&&&&&&&\\
\cline{3-10}
&&&&&&&&&\\
 $50$  & $8$ & 0.861 & 0.670 & $0.519\times 10^{15}$ & $0.936\times 10^{6}$& $0.327\times 10^{3}$ & $0.242\times 10^{1}$ &$0.159\times 10^{1}$ &0.867\\
 &&&&&&&&&\\
\cline{3-10}
&&&&&&&&&\\
   & $\infty$ &0.746 & 0.657 & $0.697\times 10^{17}$ & $0.139\times 10^{8}$ & $0.104\times 10^{3}$ &$ 0.149 \times 10^{1}$ & $0.144\times 10^{1}$&0.837\\
   &&&&&&&&&\\
\hline
&&&&&&&&&\\
 & $2$ & 0.688 & 0.648 & $0.112\times 10^{16}$ & $0.223\times 10^{7}$ & $0.609\times 10^{6}$& $0.458\times 10^{2}$ &0.986 &0.724\\
 &&&&&&&&&\\
\cline{3-10}
&&&&&&&&&\\
 $100$  & $8$ & 0.561 & 0.588 & $0.951\times10^{13}$ & $0.119 \times 10^{6}$& $0.142\times 10^{7}$ & $0.524\times 10^{2}$ &$0.123 \times 10^{1}$&0.794\\
 &&&&&&&&&\\
\cline{3-10}
&&&&&&&&&\\
   & $\infty$ & 0.554 & 0.587 & $0.968\times 10^{11}$ & $0.114 \times 10^{5}$& $0.920 \times 10^{5}$& $0.147\times 10^{2}$&$0.117\times 10^{1}$ &0.771 \\
   &&&&&&&&&\\
\hline
\end{tabular}}
\caption{Mean integrated squared error and mean integrated absolute error for various combinations of sample size ($n$) and signal-to-noise ratio ($\operatorname{StN}$) for the simulation Case 1. The compared four methods are: FRRK (AIC) (proposed), PFFR (Ivanescu {\it et al.} (2015)), and PACE (AIC) and PACE (BIC) (Yao {\it et al.} (2005b)). }
\label{tab1}
\end{table}
\begin{table}[h!]
\centering 
\vspace{-1cm}
{\footnotesize\begin{tabular}{|cc|cc|cc|cc|cc|}
\hline
 & & \multicolumn{2}{c}{FRRK} & 
\multicolumn{2}{c}{PFFR} & \multicolumn{2}{c|}{PACE (AIC)}&\multicolumn{2}{c|}{PACE (BIC)}\\
\cline{3-10}
$n$ & $\operatorname{StN}$ & MISE & MIAE & MISE & MIAE &MISE & MIAE &MISE &MIAE \\
\hline
&&&&&&&&&\\
 & $2$ & $0.313 \times 10^{2}$& $0.109 \times 10^{1}$& $0.881\times 10^{15}$ & $0.216 \times 10^{7}$& $0.419\times 10^{6}$ & $0.263\times 10^{2}$ &0.758 &0.638 \\
&&&&&&&&&\\
\cline{3-10}
&&&&&&&&&\\
 $25$  & $8$ & $0.155\times 10^{2}$ & 0.879 & $0.915\times 10^{16}$ &$0.455\times 10^{7}$ & $0.256\times 10^{1}$ & $0.102\times 10^{1}$ & 0.613&0.591 \\
   &&&&&&&&&\\
\cline{3-10}
&&&&&&&&&\\
   & $\infty$ & $0.358\times 10^{1}$ & 0.586 & $0.406\times 10^{13}$ & $0.713\times 10^{5}$ & $0.277 \times 10^{1}$& $0.102\times 10^{1}$&0.612 & 0.589\\
   &&&&&&&&&\\
\hline
&&&&&&&&&\\
 & $2$ & 0.375 & 0.436 &$ 0.259\times 10^{16}$ & $0.306\times 10^{7}$ & $0.129\times 10^{3} $&$0.114\times 10^{1}$  &0.637 & 0.605\\
 &&&&&&&&&\\
\cline{3-10}
&&&&&&&&&\\
 $50$  & $8$ & 0.284 & 0.380 & $0.312\times 10^{15}$ & $0.937\times 10^{6}$ & $0.596\times 10^{1}$ & 0.968 &0.590 &0.587 \\
 &&&&&&&&&\\
\cline{3-10}
&&&&&&&&&\\
   & $\infty$ & 0.269 & 0.378 & $0.257\times 10^{14}$ & $0.192\times 10^{6}$ & $0.669\times 10^{1}$ & 0.919 &0.554 &0.570 \\
&&&&&&&&&\\
\hline
&&&&&&&&&\\
 & $2$ & 0.210 & 0.350 & $0.119\times 10^{15}$ & $0.650\times 10^{6}$ & $0.563\times 10^{4}$ &$0.470\times 10^{1}$ & 0.547&0.553 \\
 &&&&&&&&&\\
\cline{3-10}
&&&&&&&&&\\
  $100$ & $8$ & 0.171 & 0.326 & $0.317\times 10^{16}$ & $0.203\times 10^{7}$ & $0.341\times 10^{5}$& $0.871\times 10^{1}$&0.646 &0.605 \\
  &&&&&&&&&\\
\cline{3-10}
&&&&&&&&&\\
   & $\infty$ & 0.176 & 0.323 &$0.270\times 10^{15}$ & $0.118\times 10^{7}$ &$0.461\times 10^{4}$  &$0.390\times 10^{1}$ &0.632&0.592 \\
   &&&&&&&&&\\
\hline
\end{tabular}}
\caption{Mean integrated squared error and mean integrated absolute error for various combinations of sample size ($n$) and signal-to-noise ratio ($\operatorname{StN}$) for the simulation Case 2. The compared four methods are: FRRK (AIC) (proposed), PFFR (Ivanescu {\it et al.} (2015)), and PACE (AIC) and PACE (BIC) (Yao {\it et al.} (2005b)). }
\label{tab2}
\end{table}

\begin{table}[h!]
\centering 
{\footnotesize\begin{tabular}{|cc|cc|cc|cc|cc|}
\hline
 & & \multicolumn{2}{c}{FRRK} & 
\multicolumn{2}{c}{PFFR} & \multicolumn{2}{c}{PACE(AIC)}&\multicolumn{2}{c|}{PACE(BIC)}\\
\cline{3-10}
$n$ & $\operatorname{StN}$ & MISE & MIAE & MISE & MIAE &MISE & MIAE &MISE &MIAE \\
\hline
&&&&&&&&&\\
 & $2$ &1.540 &0.892 &3.116& 1.503& 10.206& 1.941 &4.095& 1.684
  \\
  &&&&&&&&&\\
\cline{3-10}
&&&&&&&&&\\
 $25$  & $8$ & 1.172 &0.776 &4.540& 1.577 & 2.200& 0.917 &1.494& 0.873
  \\
  &&&&&&&&&\\
\cline{3-10}
&&&&&&&&&\\
   & $\infty$ & 1.190 & 0.787& 3.120 &0.438&  2.734 &0.999 &1.532& 0.878 \\
   &&&&&&&&&\\
\hline
&&&&&&&&&\\
& $2$ &        0.801 &0.687 &5.376 &1.619 & 1.633 &0.873 &1.157& 0.798 \\
&&&&&&&&&\\
\cline{3-10}
&&&&&&&&&\\
$50$    & $8$ &       0.634 &0.616& 0.179& 0.131 & 1.77& 0.800& 1.253& 0.802 \\
   &&&&&&&&&\\
\cline{3-10}
&&&&&&&&&\\
   & $\infty$ &    0.636& 0.617& 1.203& 0.287&  1.300& 0.759 &1.181& 0.800 \\
   &&&&&&&&&\\
\hline
&&&&&&&&&\\
 & $2$ & 0.611& 0.616 &2.829 &1.411&  0.896 &0.691 &0.887 &0.693 \\
&&&&&&&&&\\
\cline{3-10}
&&&&&&&&&\\
 $100$  & $8$ &0.527& 0.570& 2.753&1.400 & 0.953 &0.649 &1.033 &0.743 \\
   &&&&&&&&&\\
\cline{3-10}
&&&&&&&&&\\
   & $\infty$ & 0.519 &0.570& 0.016& 0.049&  1.047 &0.676& 0.991& 0.725 \\
   &&&&&&&&&\\
\hline
\end{tabular}}
\caption{The values in this table correspond to Table \ref{tab1} and are the 5\% upper trimmed mean integrated squared error and mean integrated absolute error for various combinations of sample size ($n$) and signal-to-noise ratio ($\operatorname{StN}$) for the simulation Case 1 . The compared four methods are: FRRK (AIC) (proposed), PFFR (Ivanescu {\it et al.} (2015)), and PACE (AIC) and PACE (BIC) (Yao {\it et al.} (2005b)). }
\label{tab3}
\end{table}
 
\begin{table}[h!]
\centering 
{\footnotesize\begin{tabular}{|cc|cc|cc|cc|cc|}
\hline
 & & \multicolumn{2}{c}{FRRK} & 
\multicolumn{2}{c}{PFFR} & \multicolumn{2}{c}{PACE(AIC)}&\multicolumn{2}{c|}{PACE(BIC)}\\
\cline{3-10}
$n$ & $\operatorname{StN}$ & MISE & MIAE & MISE & MIAE &MISE & MIAE &MISE &MIAE \\
\hline
&&&&&&&&&\\
 & $2$ &0.506& 0.495& 6.200 &1.403& 0.839& 0.648& 0.597& 0.591\\
&&&&&&&&&\\
\cline{3-10}
&&&&&&&&&\\
  $25$ & $8$ &0.315& 0.418 &2.162& 1.147& 1.497& 0.923 &0.557& 0.567 \\
   &&&&&&&&&\\
\cline{3-10}
&&&&&&&&&\\
   & $\infty$ &0.291 &0.405& 1.586& 1.075& 1.552 &0.930& 0.554&0.564 \\
   &&&&&&&&&\\
\hline
&&&&&&&&&\\
 & $2$& 0.255 &0.385& 1.672&1.078& 0.881& 0.700& 0.554& 0.573 \\
&&&&&&&&&\\
\cline{3-10}
&&&&&&&&&\\
  $50$ & $8$ &0.1974& 0.343 &1.218& 1.010&1.0523& 0.769& 0.530 &0.561\\
   &&&&&&&&&\\
\cline{3-10}
&&&&&&&&&\\
   & $\infty$& 0.200& 0.344& 1.217&1.011 &1.028& 0.759& 0.496 &0.545 \\
   &&&&&&&&&\\
\hline
&&&&&&&&&\\
 & $2$& 0.173& 0.326 &1.230 &1.015& 0.464& 0.502& 0.462& 0.519 \\
&&&&&&&&&\\
\cline{3-10}
&&&&&&&&&\\
$100$   & $8$ &0.151& 0.310& 1.188& 1.006& 0.518 &0.533& 0.558& 0.570 \\
   &&&&&&&&&\\
\cline{3-10}
&&&&&&&&&\\
   & $\infty$ &0.147& 0.304 &1.185& 1.006& 0.490& 0.518& 0.526& 0.551 \\
   &&&&&&&&&\\
\hline
\end{tabular}}
\caption{The values in this table correspond to Table \ref{tab2} and are the 5\% upper trimmed mean integrated squared error and mean integrated absolute error for various combinations of sample size ($n$) and signal-to-noise ratio ($\operatorname{StN}$) for the simulation Case 2 . The compared four methods are: FRRK (AIC) (proposed), PFFR (Ivanescu {\it et al.} (2015)), and PACE (AIC) and PACE (BIC) (Yao {\it et al.} (2005b)).}
\label{tab4}
\end{table}

\subsection{Application}
It is a well known fact that the human immune deficiency virus (HIV) causes AIDS by attacking immune cells called CD4+. A healthy person has around 1100 CD4+ cells per cubic millimetre of blood. The CD4+ cell counts can vary day to day, and even from  time to time over a day. As the number of CD4+ cells decreases, the immune system becomes weaker and therefore the likelihood of an opportunistic infection increases.  An HIV infected person's CD4+ cell counts over the time is normally used to monitor AIDS progression. It is somewhat believed that depressive symptoms are negatively correlated with the capacity for immune system response. So it is interesting to check whether CD4+ cell counts are associated with the depressed mood of an individual over time. One of the most common measures of depressive symptoms is the CES-D scale. The CES-D scale is a short self-report measure of depressive feelings and behaviours of an individual during the past week. The higher the CES-D score, the greater the depressive symptoms. 

The dataset that we use in this article comes from the well known multicenter AIDS Cohort Study. See Kaslow {\it et al.} (1987) and Diggle {\it et al.} (2002). In this study, the CD4+ cell counts and the CES-D scores of the patients  were scheduled to be measured and recorded twice a year, but because of missing appointments and other factors the actual times of measurements are random and often sparse. This dataset consists of 2376 values of CD4+ cell counts, CES-D scores and other variables over time for 369 infected men enrolled in the study which accounts to 1 to 12 observations per patient. 

\begin{figure}
\centering
\includegraphics[scale=0.35]{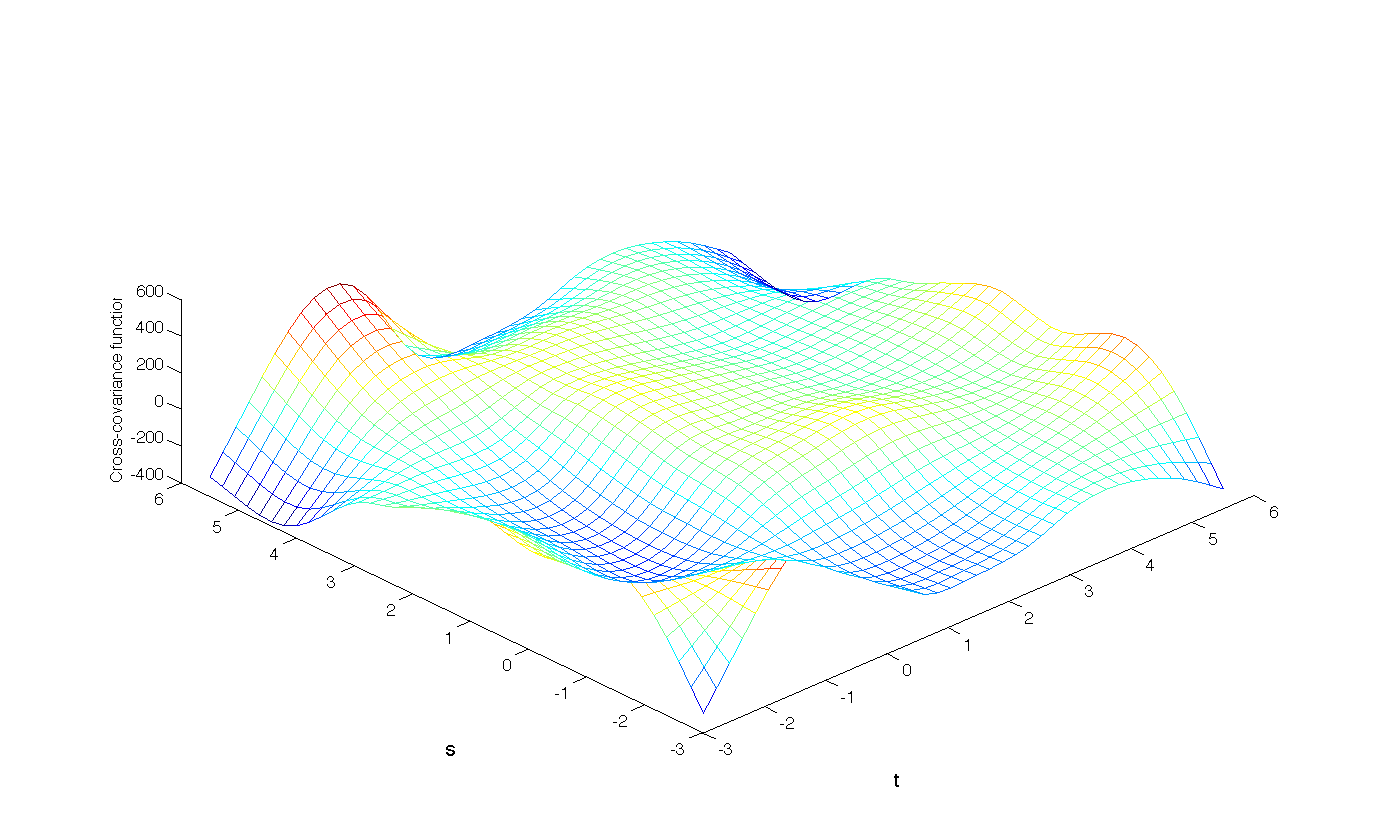}
\caption{Smoothed cross covariance surface of CES-D scores and CD4+ counts.} \label{CrossFig}
\addcontentsline{lof}{section}{
Cross
}
\end{figure}

To explore the association between the longitudinally measured response process CD4+ counts and the predictor process CES-D scores, we apply the regularization method proposed in this article. We apply a 5-fold cross validation to select the tuning parameter $\lambda$. In this dataset, both the CD4+ counts and CES-D scores are considered functions of time since seroconversion. The estimated cross covariance function surface is given in Figure \ref{CrossFig} and displays many peaks and valleys. This cross covariance surface indicates that the correlation between CD4+ count and the CES-D score processes seems to be negligible after seroconversion. For other time periods, the association seems to be mainly negative. 

\begin{figure}
\centering
\includegraphics[scale=0.75]{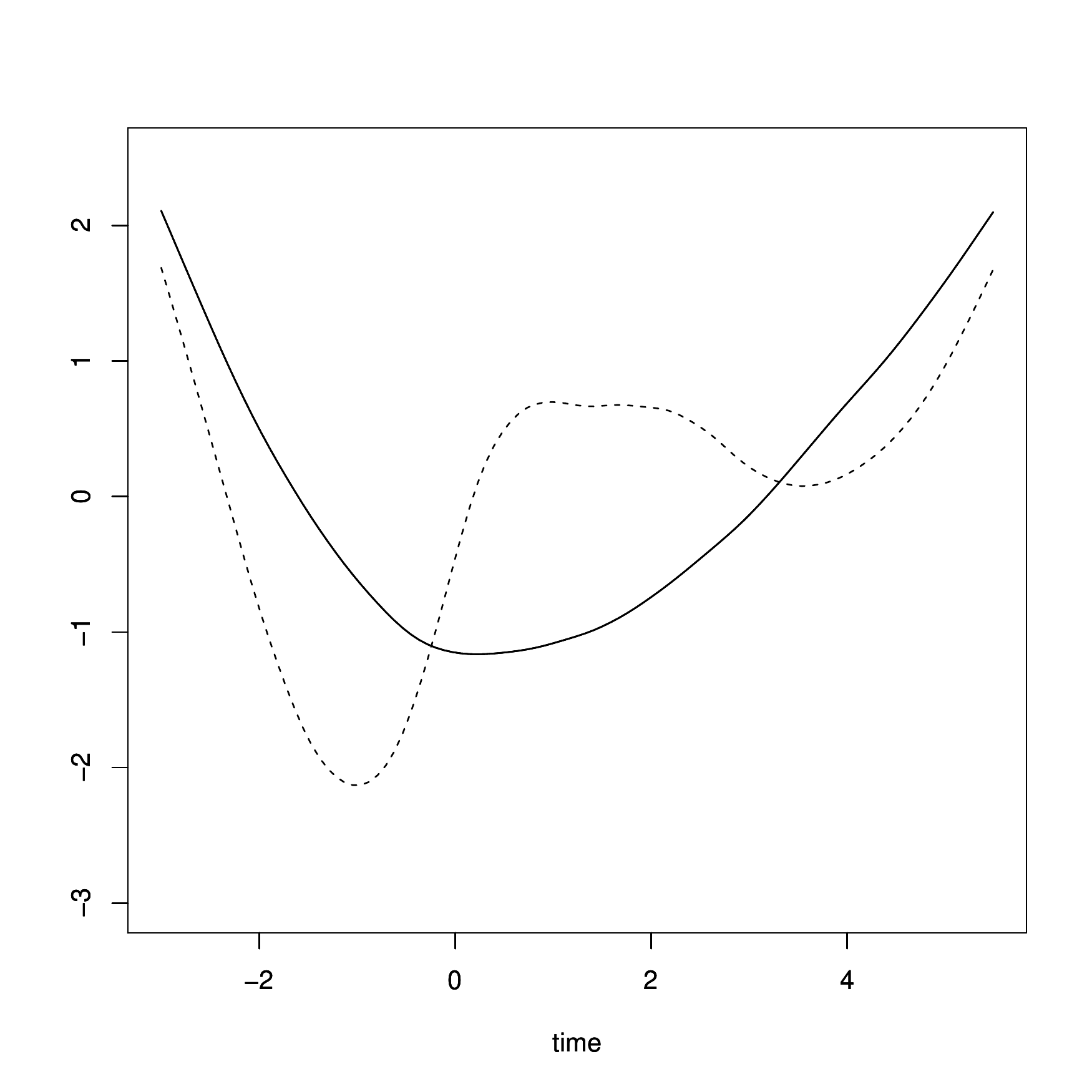}
\caption{Smooth estimate of the first pair of singular functions of CES-D scores (solid) and CD4+ counts (dashed).}\label{SingularFig}
\addcontentsline{lof}{section}{
Singular
}
\end{figure}

The estimated first pair of singular functions ($\widehat{\psi}_{_1},\,\widehat{\phi}_{_1}$) for CES-D and CD4+, respectively, from the cross-covariance function are displayed in Figure \ref{SingularFig}. The singular function $\widehat{\psi}_{_1}$ initially decreases rapidly and then increases with a slower rate. In comparison, prior to time -1, the singular function $\widehat{\phi}_{_1}$ decreases rapidly and then increases with similar rate until time 1 which stays relatively constant and then from time 4 rapidly increases again. In summary the behaviour of $\widehat{\psi}_{_1}$ and $\widehat{\phi}_{_1}$ is similar prior to time -1 and after time 4. In addition, this figure suggests that the association between the CES-D and the CD4+ is negligible for times between $0$ and $4$. It also indicates that for times outside the interval $[0,\,4]$, there is a non-negligible correlation between the CES-D and the CD4+, although not so salient.

\begin{figure}
\centering
\includegraphics[scale=0.7]{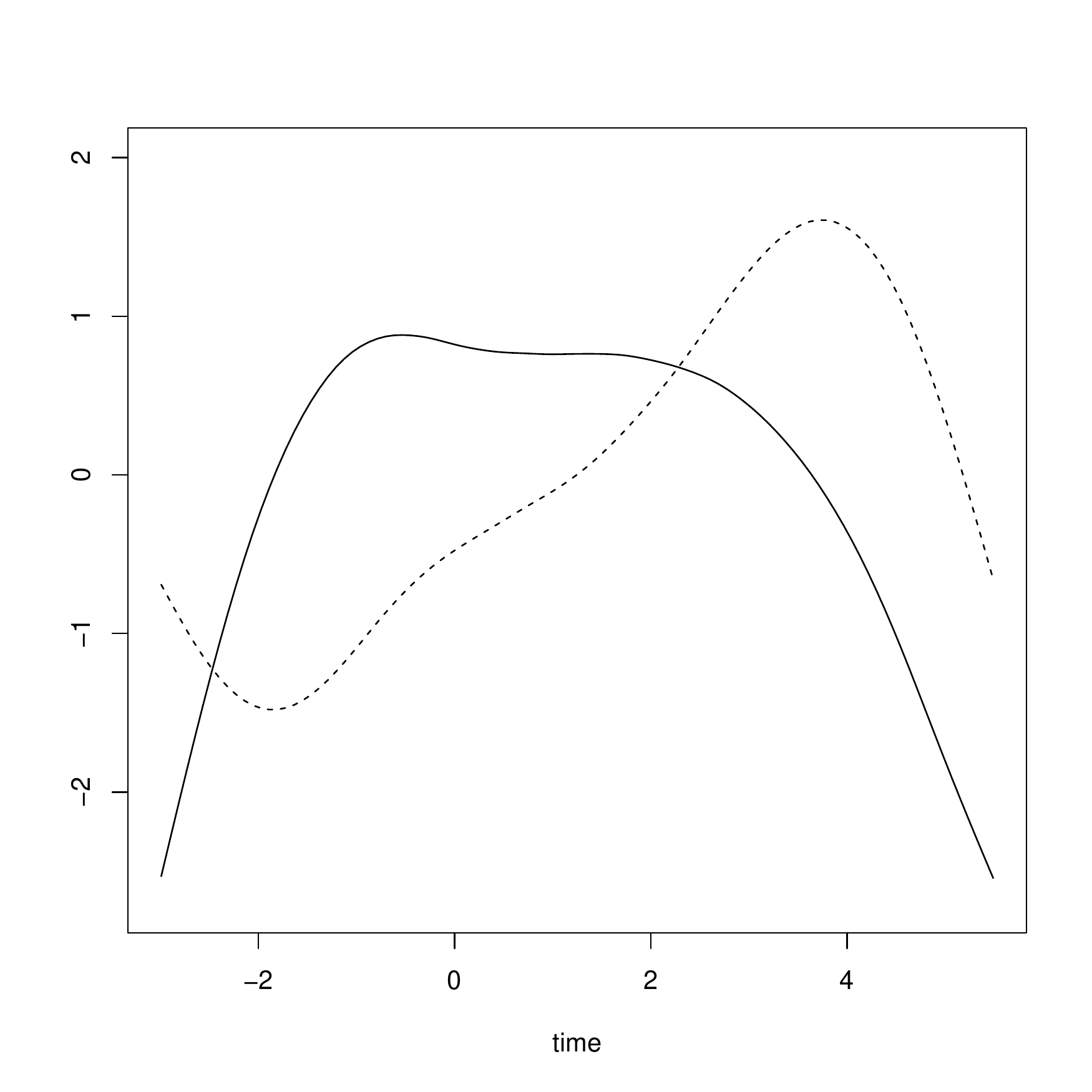}
\caption{Smooth estimate of the first (solid) and the second (dashed) eigenfunctions for CES-D scores, according to 60 and 39 percent of total variation.}\label{eigenX}
\addcontentsline{lof}{section}{
eigenX
}
\end{figure}

\begin{figure}
\centering
\includegraphics[scale=0.7]{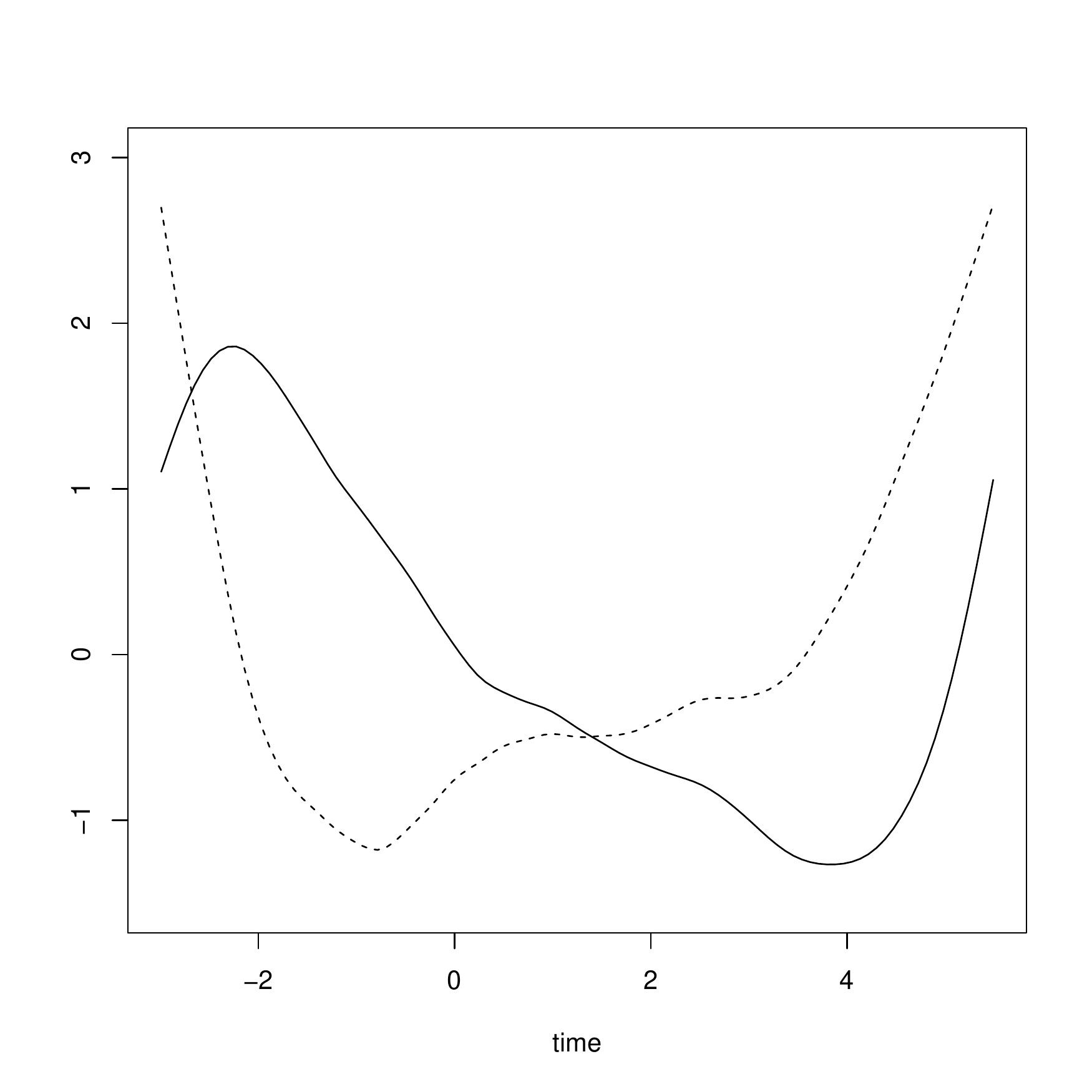}
\caption{Smooth estimate of the first (solid) and the second (dashed) eigenfunctions for CD4+ counts, according to 54 and 45 percent of total variation.}\label{eigenY}
\addcontentsline{lof}{section}{
eigenY
}
\end{figure}

The first and the second eigenfunctions of CES-D and CD4+ are shown in Figures \ref{eigenX} and \ref{eigenY}, respectively. The first eigenfunction of CES-D corresponds to a contrast between midtime (which is relatively constant) and early and late times (which is an indicative of rapid increasing and decreasing behaviour). The second eigenfunction of CES-D has an oscillative behaviour.  On the other hand, the first eigenfunction of CD4+ has an oscillative behaviour and the second eigenfunction of CD4+ corresponds to contrast between midtime and early and late times. 

\begin{figure}
\centering
\includegraphics[scale=0.35 ]{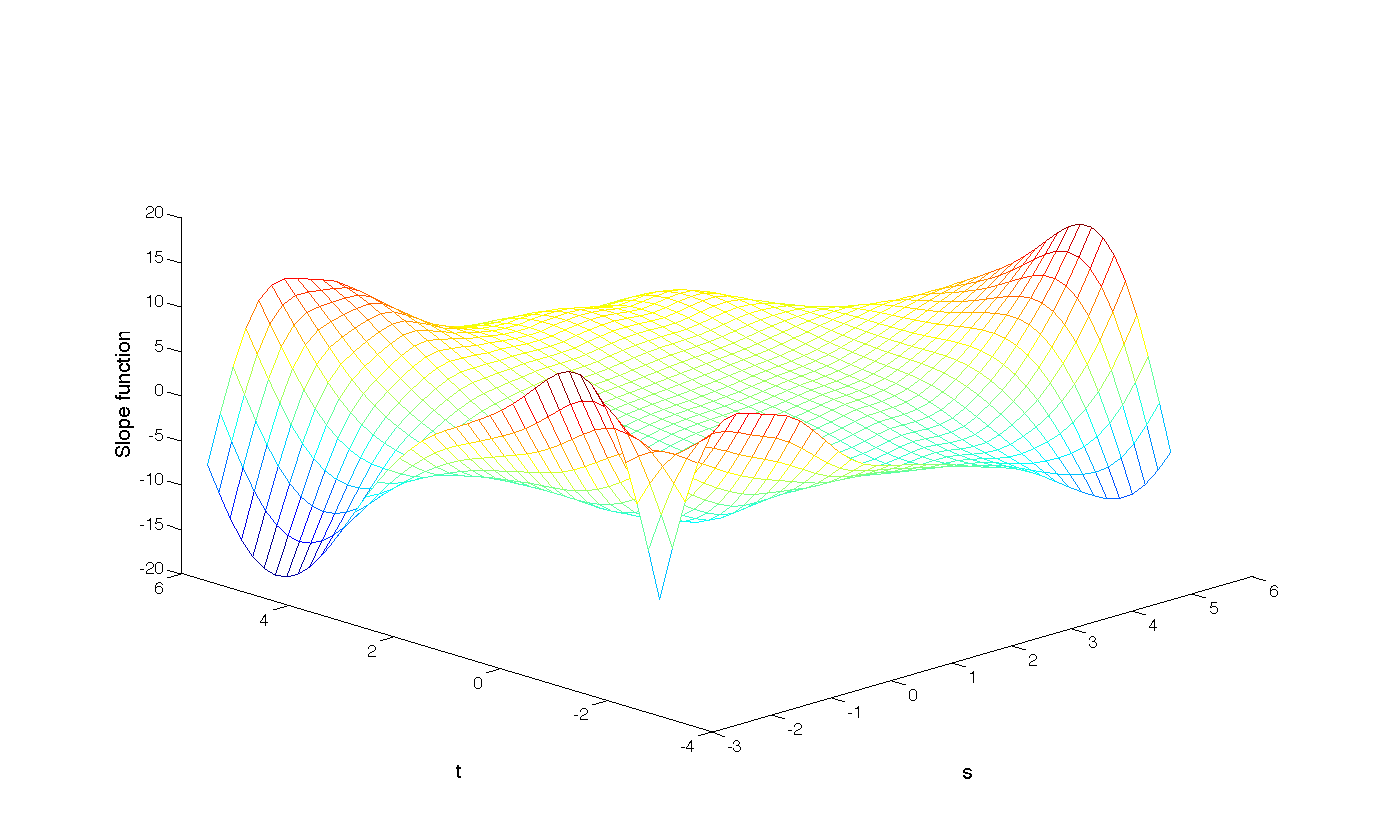}
\caption{Estimated slope coefficient function \eqref{slophat}, where the predictor (CES-D scores) time is $s$, and the response (CD4+ counts) time is $t$.} \label{SlopFig}
\addcontentsline{lof}{section}{
slope
}
\end{figure}

The estimate of the coefficient function $\beta$ is given in Figure \ref{SlopFig}. Recall that for the functional linear regression \eqref{flr1.1} studied in this article, the coefficient function $\beta(s,\,t)$ can be interpreted as the relative weight placed on $X(s)$ at time $s$ which is required to predict $Y(t)$ on a fixed time $t$. Therefore, the shape of $\widehat{\beta}(s,t)$ in Figure \ref{SlopFig} indicates that, except for the time points near the boundary, the contribution of the CES-D on the CD4+ is negligible. In the corners, we observe minor association. This association may be influenced by very high sparsity of the data at those times. For example only $2.9 \%$ of time points are greater than $4.5$ while the length of this sub-interval is $11.3 \%$ of that for $\mathcal{T}$. Overall, taking into account the sparsity of the data, there is not any noticeable association.

\newpage
\section{Proofs}
\subsection{Proof of Theorem 1}
Proof of $C_{_{X\,Y}},\,C_{_{Y\,X}}\in \mathcal{H}(K\otimes K)$ is similar to that of $C_{_{X\,X}}\in \mathcal{H}(K\otimes K)$ in Cai and Yuan (2010) and therefore it is omitted. We only prove $A_{_{X\,Y}}\in \mathcal{H}(K\otimes K)$, as $A_{_{Y\,X}}\in \mathcal{H}(K\otimes K)$ is similar. First note that 
\begin{align*}
 A_{_{X\,Y}}(s,\,t) &= \int_\mathcal{T} C_{_{X\,Y}}(s,\,u)\,C_{_{X\,Y}}(t,\,u)\,du  \\
& = \int_\mathcal{T} \left\lbrace E[\,X(s)\,Y(u)\,] - \mu_{_X}(s)\,\mu_{_Y}(u)\right\rbrace\,\left\lbrace E[\,X(t)\,Y(u)\,] - \mu_{_X}(t)\,\mu_{_Y}(u)\right\rbrace\,du\\
&= \underbrace{\int_\mathcal{T} E[\,X(s)\,Y(u)\,]\,E[\,X(t)\,Y(u)\,]\,du}_I + \underbrace{\mu_{_X} (s)\,\mu_{_X}(t)\int_\mathcal{T} \mu_{_Y}^2 (u)\, du}_{II} \\
& - \underbrace{\mu_{_X}(t)\,\int_\mathcal{T} E[\,X(s)\,Y(u)\,]\,\mu_{_Y}(u)\,du}_{III} - \underbrace{\mu_{_X}(s)\,\int_\mathcal{T}E[\,X(t)\,Y(u)\,]\,\mu_{_Y} (u)\,du}_{IV}\,. 
\end{align*} 

\noindent
Therefore, it suffices to show that $I,\,II,\,III,\,IV\in \mathcal{H}(K\otimes K)$.
From \eqref{RXY}, we have
\begin{align*}
  I &= \int_\mathcal{T} \left\lbrace\sum\limits_{j,\, k \geq 1} E[\,x_{_j}\, y_{_k}\,]\,\varphi_{_j}(s)\,\varphi_{_k}(u)\right\rbrace\,\left\lbrace\sum\limits_{j',\,k' \geq 1} E[\,x_{_{j'}} \,y_{_{k'}}\,]\,\varphi_{_{j'}}(t)\,\varphi_{_{k'}}(u)\right\rbrace \,du   \\
 & = \int_\mathcal{T} \sum\limits_{j,\,k \geq 1}\, \sum\limits_{j',\,k' \geq 1} \left\lbrace E[\,x_{_j} \,y_{_k}\,] \,E[\,x_{_{j'}} \,y_{_{k'}}\,]\,\varphi_{_j}(s)\,\varphi_{_k}(u)\,\varphi_{_{j'}}(t)\,\varphi_{_{k'}}(u) \right\rbrace\, du  \\
 & = \sum\limits_{j,\,j' \geq 1} \left\lbrace  \sum\limits_{k,\,k' \geq 1} E[\,x_{_j} \,y_{_k} \,]\,E[\,x_{_{j'}} \,y_{_{k'}} \,]\,\int_\mathcal{T}\varphi_{_k}(u)\,\varphi_{_{k'}}(u)\, du \right\rbrace\,\varphi_{_j}(s)\,\varphi_{_{j'}}(t)\,.
\end{align*} 
Since $\lbrace\varphi_{_k} ;\,k \geq 1\rbrace$ is an orthonormal basis of $L_2(\mathcal{T})$, we have
$$
I= \sum\limits_{j,\,j' \geq 1}h_{_{j\,j'}}\, \varphi_{_j} (s)\,\varphi_{_{j'}} (t)\,,
$$
where
$$
h_{_{j\,j'}}  =\sum\limits_{k \geq 1} E[\,x_{_j} \,y_{_k} \,]\,E[\,x_{_{j'}}\, y_{_k}\,]\,.
$$
Now by \eqref{Wahba}, it suffices to show that $\|\,I\,\|^2_{_{\mathcal{H}(K\otimes K)}}=\sum\limits_{j,\,j' \geq 1} \rho_{_j}^{ - 1} \,\rho_{_{j'}}^{ - 1} \,h_{_{j\,j'}}^2 <\infty$. So by applying the Cauchy-Schwarz inequality twice, we obtain
\begin{align*}
  h_{_{j\,j'}}^2 & \leq \left\lbrace \sum\limits_{k \geq 1} \left(E[\,x_{_j} \,y_{_k}\,]\right)^2 \right\rbrace\,\left\lbrace \sum\limits_{k' \geq 1} \left(E[\,x_{_{j'}} \,y_{_{k'}}\,]\right)^2 \right\rbrace \\
 & \leq \left\lbrace \sum\limits_{k \geq 1} E[\,x_{_j}^2 \,]\,E[\,y_{_k}^2\, ] \right\rbrace\,\left\lbrace \sum\limits_{k' \geq 1} E[\,x_{_{j'}}^2\,]\,E[\,y_{_{k'}}^2\,] \right\rbrace\\
  & = \left\lbrace E[\,x_{_j}^2\,]\,E[\,x_{_{j'}}^2\,] \right\rbrace\,\left\lbrace E\|\,Y\,\|_{_{L_2}}^2 \right\rbrace^2\,.
\end{align*} 

\noindent
Therefore

\begin{align*}
  \sum\limits_{j,\,j' \geq 1} \rho_{_j}^{ - 1} \rho_{_{j'}}^{ - 1} h_{_{j\,j'}}^2 & \leq \left\lbrace E\|\,Y\,\|_{_{L_2 }}^2 \right\rbrace^2 \sum\limits_{j,\,j' \geq 1} \rho_{_j}^{ - 1} \rho_{_{j'}}^{ - 1}\, E[\,x_{_j}^2\,]\,E[\,x_{_{j'}}^2\,] \\
 &= \left\lbrace E\|\,Y\,\|_{_{L_2}}^2\right\rbrace^2\, \left\lbrace E\|\,X\,\|_{_{\mathcal{H}(K)}}^2 \right\rbrace^2  < \infty  
\end{align*}
which implies that $I\in \mathcal{H}(K\otimes K)$.\\

\noindent
Since $$
\int_\mathcal{T} \mu_{_Y}^2 (u)\,du =\|\,E\left[Y\right]\|_{_{L_2 }}^2  \leq E\|\,Y\,\|_{_{L_2 }}^2  < \infty 
$$
then $II\in \mathcal{H}(K\otimes K)$.\\

\noindent
To prove $III\in \mathcal{H}(K\otimes K)$, we first note that 
\begin{align*}
  \int_\mathcal{T} E\left[\,X(s)\,Y(u)\,\right]\,\mu_{_Y}(u)\,du & = \int_\mathcal{T}\left\lbrace\sum\limits_{j,\,k \geq 1} E[\,x_{_j}\, y_{_k}\,]\,\varphi_{_j}(s)\,\varphi_{_k}(u)\right\rbrace\,\left\lbrace\sum\limits_{\ell \geq 1} E[\,y_{_\ell}\,]\,\varphi_{_\ell} (u)\right\rbrace\, du\\
  & = \sum\limits_{j,\,k,\,\ell \geq 1} E[\,x_{_j}\, y_{_k}\,]\,E[\,y_{_\ell}\,]\,\varphi_{_j}(s)\,\int_\mathcal{T}\varphi_{_k}(u)\,\varphi_{_\ell}(u)\,du\\
  & = \sum\limits_{j,\,k \geq 1} E[\,x_{_j} \,y_{_k}\,]\,E[\,y_{_k}\,]\,\varphi_{_j}(s) \\
  &= \sum\limits_{j \geq 1}g_{_j}\, \varphi_{_j}(s) \,,
\end{align*} 
where 
$$
g_{_j}=\sum\limits_{k \geq 1} E[\,x_{_j} \,y_{_k}\,]\,E[\,y_{_k}\,]\,. 
$$
Thus by twice application of the Cauchy-Schwarz inequality
\begin{align*}
g_{_j}^2&=\left(\sum\limits_{k \geq 1} E[\,x_{_j} \,y_{_k}\,]\,E[\,y_{_k}\,]\right)^2 \\
& \leq \left\lbrace\sum\limits_{k \geq 1} \left(E[\,x_{_j} \,y_{_k}\,]\right)^2 \right\rbrace\, \left\lbrace \sum\limits_{k \geq 1} \left(E[\,y_{_k}\,] \right)^2 \right\rbrace \\
&\leq \left\lbrace \sum\limits_{k \geq 1} E[\,x_{_j}^2\,]\,E[\,y_{_k}^2 \,] \right\rbrace\, \left\lbrace \sum\limits_{k \geq 1} E[\,y_{_k}^2 \,]  \right\rbrace \\
& = E[\,x_{_j}^2\,]\,\left\lbrace E\|\,Y\,\|_{_{L_2}}^2 \right\rbrace^2\,.
\end{align*} 

\noindent 
Therefore,
$$
\sum\limits_{j \geq 1} \rho_{_j}^{-1}\, g_{_j}^2  \leq \left\lbrace E\|\,X\,\|_{_{\mathcal{H}(K)}}^2 \right\rbrace\,\left\lbrace E\|\,Y\,\|_{_{L_2}}^2 \right\rbrace^2 < \infty, 
$$
which implies $\int_\mathcal{T} E\left[\,X(s)\,Y(u)\,\right]\,\mu_{_Y}(u)\,du \in \mathcal{H}(K)$, that is $III\in \mathcal{H}(K\otimes K)$.  Similarly, we can show that $IV\in \mathcal{H}(K\otimes K)$. $\blacksquare$

\subsection{Proof of Lemma 1}
For brevity in notation, let $\mathcal{H}:=\mathcal{H}(K \otimes K)$ and $\widehat C:=\widehat C_{_{X\,Y}}$. Set $D=C-\widehat C$. Then,
\begin{equation*}
\|\,C\,\|_{_\mathcal{H}}^2  = \|\,D + \widehat{C}\,\|_{_\mathcal{H}}^2  = \|\,D\,\|_{_\mathcal{H}}^2  + 2\, \langle\, \widehat{C},\,D \,\rangle_{_\mathcal{H}} + \|\,\widehat{C}\|_{_\mathcal{H}}^2\,. 
\end{equation*}
Also
\begin{align*}
 \ell_n(C) &= \frac{1}
{n\,m_{_1} \,m_{_2}}\sum\limits_{i = 1}^n \sum\limits_{j_{_1}  = 1}^{m_{_1}} \sum\limits_{j_{_2}  = 1}^{m_{_2} } \left[\, C_i(S_{i\,j_{_1}} ,\,T_{i\,j_{_2}}) - \widehat{C}(S_{i\,j_{_1}} ,\,T_{i\,j_{_2}}) - D(S_{i\,j_{_1}} ,\,T_{i\,j_{_2}})\,\right]^2  \\
 & = \frac{1}
{n\,m_{_1} \,m_{_2}}\left\lbrace \sum\limits_{i = 1}^n \,\sum\limits_{j_{_1} = 1}^{m_{_1}} \,\sum\limits_{j_{_2} = 1}^{m_{_2}} \left[\,C_i(S_{i\,j_{_1} } ,\,T_{i\,j_{_2}}) - \widehat{C}(S_{ij_{_1}} ,\,T_{i\,j_{_2}})\,\right]^2 + \sum\limits_{i = 1}^n \sum\limits_{j_{_1}= 1}^{m_{_1}} \sum\limits_{j_{_2}= 1}^{m_{_2}} D^2(S_{i\,j_{_1}} ,\,T_{i\,j_{_2}}) \right.\\
&\left.- 2\,\sum\limits_{i = 1}^n \sum\limits_{j_{_1}= 1}^{m_{_1}} \sum\limits_{j_{_2} = 1}^{m_{_2}} \left[\,C_i (S_{i\,j_{_1}} ,\,T_{i\,j_{_2}}) - \widehat{C}(S_{i\,j_{_1}} ,\,T_{i\,j_{_2}} )\,\right]\,D(S_{i\,j_{_1}} ,\,T_{i\,j_{_2} } ) \right\rbrace\,.
\end{align*} 
Therefore
\begin{align*}
E(C) &= \ell _n (C) + \lambda\, \|\,C\,\|_{_\mathcal{H}}^2 \\
  &= \ell _n (\widehat{C}) + \lambda\, \|\,\widehat{C}\,\|_{_\mathcal{H}}^2  + \lambda\,\|\,D\,\|_{_\mathcal{H}}^2  + \frac{1}
{n\,m_{_1}\, m_{_2} }\sum\limits_{i = 1}^n \sum\limits_{j_{_1}  = 1}^{m_{_1} } \sum\limits_{j_{_2}  = 1}^{m_{_2} } D^2 (S_{i\,j_{_1} } ,\,T_{i\,j_{_2}} ) \\
&+ 2\,\lambda\, \langle\, \widehat{C},\,D \,\rangle _{_\mathcal{H} } -\, \frac{2}
{n\,m_{_1}\, m_{_2} }\sum\limits_{i = 1}^n \sum\limits_{j_{_1}  = 1}^{m_{_1} } \sum\limits_{j_{_2}  = 1}^{m_{_2} } \left[C_i (S_{i\,j_{_1} } ,\,T_{i\,j_{_2} } ) - \widehat{C}(S_{i\,j_{_1} } ,\,T_{i\,j_2 } )\right]\,D(S_{i\,j_{_1} } ,\,T_{i\,j_{_2} } )
\end{align*} 
and consequently
\begin{align}\label{E(C)2}
  E(C)& = E(\widehat{C}) + \lambda\, \|\,D\,\|_{_\mathcal{H}}^2  + \frac{1}
{n\,m_{_1} \,m_{_2}}\sum\limits_{i = 1}^n \sum\limits_{j_{_1} = 1}^{m_{_1}} \sum\limits_{j_{_2}  = 1}^{m_{_2} } D^2 (S_{i\,j_{_1} } ,\,T_{i\,j_{_2} } )  \\
  &+ 2\, \lambda\,  \langle\, \widehat{C},\,D \,\rangle_{_\mathcal{H}} - \frac{2}
{n\,m_{_1}\, m_{_2}}\sum\limits_{i = 1}^n \sum\limits_{j_{_1}= 1}^{m_{_1} }\sum\limits_{j_{_2}  = 1}^{m_{_2} }\left[C_i (S_{i\,j_{_1}} ,T_{i\,j_{_2} }) - \widehat{C}(S_{i\,j_{_1}} ,\,T_{i\,j_{_2}} )\right]\,D(S_{i\,j_{_1} } ,\,T_{i\,j_{_2} } )\,. \nonumber
\end{align}
By using equation (\ref{represent}) and reproducing property of $K\otimes K$, we obtain
\begin{equation}\label{inner}
\begin{split}
  \langle \widehat{C},\,D \rangle_{_\mathcal{H}}&  =  \langle \,\sum\limits_{i = 1}^n \sum\limits_{j_{_1}  = 1}^{m_{_1} } \sum\limits_{j_{_2}  = 1}^{m_{_2} } a_{i\,j_{_1}\, j_{_2}}\, K \otimes K(\cdot,\,(S_{i\,j_{_1}} ,\,T_{i\,j_{_2} } )) ,\,D\,\rangle_{_\mathcal{H}} \\
& = \sum\limits_{i = 1}^n \sum\limits_{j_{_1}  = 1}^{m_{_1} } \sum\limits_{j_{_2}  = 1}^{m_{_2}} a_{i\,j_{_1} \,j_{_2}}\, D(S_{i\,j_{_1}} ,\,T_{i\,j_{_2}})\,. 
\end{split} 
\end{equation}
Substituting equations (\ref{subject}) and \eqref{inner} in \eqref{E(C)2} yields
\begin{align*}
E(C) &= E(\widehat{C}) + \lambda\, \|\,D\,\|_{_\mathcal{H}}^2  + \frac{1}
{n\,m_{_1}\, m_{_2} }\sum\limits_{i = 1}^n \sum\limits_{j_{_1} = 1}^{m_{_1}} \sum\limits_{j_{_2}  = 1}^{m_{_2} } D^2 (S_{i\,j_{_1}} ,\,T_{i\,j_{_2}}) \\
&\geq E(\widehat{C})\,. 
\end{align*} 
Since $\widehat{C} \in \mathcal{H}$ then $\widehat{C}=\underset{C \in \mathcal{H}}{\operatorname{argmin}}\,E(C)$. $\blacksquare$

\subsection{Proof of Lemma 2}
We first show that $\widehat \psi _{_k}$s are orthonormal. Note that 
\begin{equation*}
\int_\mathcal{T} \mathbf{g}_{_1}(s)\,\mathbf{g}'_{_1}(s)\,ds =\mathbf{P}\,.
\end{equation*}
Therefore,
$$
 \int_\mathcal{T} \widehat{ \psi}_{_{k_{_1}} }(s)\,\widehat{\psi} _{_{k_{_2}} }(s)\,ds  = \bfmath{\alpha}'_{_{k_{_1 }}}\, \mathbf{P}\,\bfmath{\alpha}_{_{k_{_2}}} = \mathbf{w}'_{_{1\,k_{_1}} } \, \mathbf{w}_{_{1\,k_{_2}} } = \delta _{_{k_{_1} \,k_{_2} }} \,, $$
where $\mathbf{w}_{_{1\,k_{_1}} }$ and $\mathbf{w}_{_{1\,k_{_2}} }$ are the $k_{_1}$th and $k_{_2}$th columns of matrix $\mathbf{W}_{_1}$. 

\noindent
Similarly for $\widehat{\phi}_{_k}$s, we have
\begin{equation}
\int_\mathcal{T} \mathbf{g}_{_2}(s)\,\mathbf{g}'_{_2}(s)\,ds=\mathbf{Q}\,.
\end{equation}
Thus,
$$
\int_\mathcal{T}\widehat{\phi}_{_{k_{_1}}}(s)\,\widehat{\phi}_{_{k_{_2}}}(s)\,ds  = \bfmath{\beta}'_{_{k_{_1}}} \,\mathbf{Q}\,\bfmath{\beta}_{_{k_{_2}}}=\mathbf{w}'_{_{2\,k_{_1}}} \, \mathbf{w}_{_{2\,k_{_2}}}=\delta _{_{k_{_1}\, k_{_2}}}\,,
$$
where $\mathbf{w}'_{_{2\,k_{_1}} }$ and $\mathbf{w}_{_{2\,k_{_2}} }$ are the $k_{_1}$th and $k_{_2}$th columns of matrix $\mathbf{W}_{_2}$. 
Now, let $\{\widehat{\sigma}_{_k}^2\} $ be the eigenvalues of 
 $\mathbf{P}^{\frac{1}{2}}\,\mathbf{A}\,\mathbf{Q}\,\mathbf{A}'\,\mathbf{P}^{\frac{1}{2}}$.
 Then $\lbrace \widehat{\sigma}_{_k}^2\rbrace$ are also the eigenvalues of 
 $\mathbf{Q}^{\frac{1}{2}}\,\mathbf{A}'\,\mathbf{P}\,\mathbf{A}\,\mathbf{Q}^{\frac{1}{2}}$ (see, e.g. Harville (1997), Theorem 21.10.1). So 

\begin{equation}\label{l}
\begin{split}
\sum\limits_{k \ge 1} \widehat{\sigma}_{_k}^2\, \widehat{\psi}_{_k}(s)\,\widehat{\psi}_{_k}(t) & =\mathbf{g}'_{_1}(s)\left(\sum\limits_{k \ge 1} \widehat{\sigma}_{_k}^2\,\bfmath{\alpha}_{_k}\,\bfmath{\alpha}'_{_k} \right)\mathbf{g}_{_1}(t)\\
& =\mathbf{g}'_{_1}(s)\,\mathbf{P}^{- \frac{1}{2}}\,\left(\sum\limits_{k \ge 1} \widehat{\sigma}_{_k}^2\,\mathbf{w}_{_{1\,k}}\,\mathbf{w}'_{_{1\,k}} \right)\,\mathbf{P}^{- \frac{1}{2}}\,\mathbf{g}_{_1}(t)\\
& = \mathbf{g}'_{_1}(s)\,\mathbf{P}^{- \frac{1}{2}}\,\left(\mathbf{P}^{\frac{1}{2}}\,\mathbf{A}\,\mathbf{Q}\,\mathbf{A}'\,\mathbf{P}^{\frac{1}{2}} \right)\,\mathbf{P}^{- \frac{1}{2}}\,\mathbf{g}_{_1}(t)\\
& = \mathbf{g}'_{_1}(s)\,\mathbf{A}\,\left(\int_\mathcal{T}\mathbf{g}_{_2}(u)\,\mathbf{g}'_{_2}(u)\,du \right)\,\mathbf{A}' \mathbf{g}_{_1}(t)\\
& = \int_\mathcal{T} \left( \mathbf{g}'_{_1}(s)\,\mathbf{A}\,\mathbf{g}_{_2}(u)\right)\,\left(\mathbf{g}'_{_1}(t)\,\mathbf{A}\,\mathbf{g}_{_2}(u) \right)'\,du \\
& = \int_\mathcal{T} \widehat{C}_{_{XY}}(s,u)\,\widehat{C}_{_{X\,Y}}(t,u)\,du \\
& = \widehat{A}_{_{X\,Y}}(s,t)\,.
\end{split}
\end{equation}
Similarly, it can be shown that
$$\sum\limits_{k \ge 1} \widehat{\sigma}_{_k}^2 \,\widehat{\phi}_{_k}(s)\,\widehat{\phi}_k(t) = \widehat{A}_{_{Y\,X}}(s,t)$$
which completes the proof. $\blacksquare$

\subsection{Proof of Corollary 1}
We only prove \eqref{AXYRC} as the proof of \eqref{AYXRC} is similar. Note that
\begin{align*}
\Big\|\,\widehat{A}_{_{X\,Y}} - A_{_{X\,Y}}\Big\|_{_{L_2 }} &= \Big\|\,\int_\mathcal{T}\widehat{C}_{_{X\,Y}}(\cdot,\,u)\,\widehat{C}_{_{X\,Y}}(*,\,u)\,du- \int_\mathcal{T}C_{_{X\,Y}}(\cdot,\,u)\,C_{_{X\,Y}} (*,\,u)\,du\, \Big\|_{_{L_2 }} \\
& \leq \Big\|\int_\mathcal{T} \left\lbrace \widehat{C}_{_{X\,Y}}(\cdot,\,u) - C_{_{X\,Y}}(\cdot,\,u)\right\rbrace\,\left\lbrace\widehat{C}_{_{X\,Y}}(*,\,u) - C_{_{X\,Y}} (*,\,u) \right\rbrace\,du\,\Big\|_{_{L_2 }} \\
& + \Big\|\int_\mathcal{T} C_{_{X\,Y}}(\cdot,\,u)\,\left\lbrace\widehat{C}_{_{X\,Y}}(*,\,u)-C_{_{X\,Y}}(*,\,u)\right\rbrace\,du\,\Big\|_{_{L_2} }\\
& + \Big\|\int_\mathcal{T} C_{_{X\,Y}}(*,\,u)\,\left\lbrace\widehat{C}_{_{X\,Y}}(\cdot ,\,u)-C_{_{X\,Y}}(\cdot,\,u)\right\rbrace\,du\,\Big\|_{_{L_2}}   \\
& \leq \Big\|\,\widehat{C}_{_{X\,Y}}  - C_{_{X\,Y}} \,\Big\|_{_{L_2}}^2  + 2\,\Big\|\,C_{_{X\,Y}} \Big\|_{_{L_2}} \,\Big\|\,\widehat{C}_{_{X\,Y}}  - C_{_{X\,Y}} \Big\|_{_{L_2 }}\,. 
\end{align*} 
where the first inequality is obtained by applying the triangle inequality and the second is obtained by applying the Cauchy-Schwarz inequality. Then \eqref{AXYRC} follows from Theorem \ref{T3}. $\blacksquare$

\subsection{Proof of Corollary 2}
From Bhatia {\it et al.} (1983), we have 
\begin{align*}
\sup\limits_{k \ge 1} \Big|\,\widehat{\sigma}_{_k}^2 - \sigma _k^2\,\Big| & \leq \Big\|\,\widehat{A}_{_{X\,Y}} - A_{_{X\,Y}}\,\Big\|_{_{L_2 }}\,,\\
\sup\limits_{k \geq 1} \,\delta_{_k}\,\Big\|\,\widehat{\phi}_{_k} - \phi_{_k}\, \Big\| &\leq 8^{1/2} \,\Big\|\,\widehat{A}_{_{X\,Y}} - A_{_{X\,Y}}\,\Big\|_{_{L_2}}\,, \\
\sup\limits_{k \geq 1} \,\delta _{_k}\,\Big\|\,\widehat{\psi}_{_k}  - \psi_{_k}\,\Big\| & \leq 8^{1/2} \,\Big\|\,\widehat{A}_{_{Y\,X}}  - A_{_{Y\,X}} \,\Big\|_{_{L_2} },
\end{align*}
where $\delta_{_k}  = \min\limits_{1 \leqslant j \leqslant k} (\sigma_{_j}^2  - \sigma_{_{j + 1}}^2 )$. Then \eqref{52} follows from Corollary \ref{Cor1} and also equations \eqref{50} and \eqref{51} follow from Corollary \ref{Cor1} and the fact that $\sigma ^2_k$ has a multiplicity one. $\blacksquare$

\subsection{Proof of Theorem 3}
We first note that 
\begin{align*}
\Big\|\,\widehat{\beta} - \beta \Big\|_{_{L_2 }}^2  &= \int_{\mathcal{T}\times \mathcal{T}} \left[\,\widehat{\beta}(s,\,t) - \beta(s,\,t)\,\right]^2\, ds\,dt \\
& = \int_{\mathcal{T}\times \mathcal{T}} \left\lbrace \sum\limits_{k = 1}^{J_{_1} - 1} \sum\limits_{\ell = 1}^{J_{_2} - 1} \left[\,\frac{\widehat{\sigma}_{_{k\,\ell}}}{\widehat{\lambda}_{_{X\,\ell}}}\,\widehat{\Psi}_{_\ell}(s)\,\widehat{\Phi}_{_k}(t)-\frac{\sigma_{_{k\,\ell}}}{\lambda_{_{X\,\ell}}}\,\Psi_{_\ell}(s)\,\Phi_{_k}(t) \right] \right\rbrace^2 \,ds\,dt\\
& + \sum\limits_{k = J_{_1}}^\infty \sum\limits_{\ell = J_{_2}}^\infty \frac{\sigma_{_{k\,\ell}}^2}{\lambda_{_{X\,\ell}}^2}  \\
& + \int_{\mathcal{T}\times \mathcal{T}}\left\lbrace\sum\limits_{k = 1}^{J_{_1} - 1}\sum\limits_{\ell = 1}^{J_{_2} - 1} \left[\,\frac{\widehat{\sigma}_{_{k\,\ell}}}{\widehat{\lambda}_{_{X\,\ell}}}\,\widehat{\Psi}_{_\ell}(s)\,\widehat{\Phi}_{_k}(t)-\frac{\sigma_{_{k\,\ell}}}{\lambda_{_{X\,\ell}}}\,\Psi_{_\ell}(s)\,\Phi_{_k}(t) \right] \right\rbrace\\
&\times \left\lbrace \sum\limits_{k = J_{_1}}^\infty \sum\limits_{\ell = J_{_2}}^\infty \frac{\sigma_{_{k\,\ell}}}
{\lambda_{_{X\,\ell}} }\,\Psi_\ell(s)\,\Phi_k (t) \right\rbrace\,ds\,dt \\
& = f_{_1}  + f_{_2}  + f_{_3}\,.  \\ 
\end{align*} 
By the Cauchy Schwartz inequality
\begin{equation*}
f_{_1}\leq \sum\limits_{k = 1}^{J_{_1} - 1} \sum\limits_{\ell = 1}^{J_{_2} - 1} \int_{\mathcal{T} \times \mathcal{T}} \left[\,\frac{\widehat{\sigma}_{_{k\,\ell}}}{\widehat{\lambda}_{_{X\,\ell}}}\,\widehat{\Psi}_{_\ell}(s)\,\widehat{\Phi}_{_k}(t)-\frac{\sigma_{_{k\,\ell}}}{\lambda_{_{X\,\ell}}}\,\Psi_{_\ell}(s)\,\Phi_{_k}(t) \right]^2 \,ds\,dt\,.\\
\end{equation*}
Also
\begin{align*}
\int_{\mathcal{T} \times \mathcal{T}} &\left[\,\frac{\widehat{\sigma}_{_{k\,\ell}}}{\widehat{\lambda}_{_{X\,\ell}}}\,\widehat{\Psi}_{_\ell}(s)\,\widehat{\Phi}_{_k}(t)-\frac{\sigma_{_{k\,\ell}}}{\lambda_{_{X\,\ell}}}\,\Psi_{_\ell}(s)\,\Phi_{_k}(t) \right]^2 \,ds\,dt\\
&\le \left[\int_{\mathcal{T} \times \mathcal{T}} \Psi_{_\ell}^2(s)\,\Phi_{_k}^2(t) \left(\frac{\widehat{\sigma}_{_{k\,l}}}{\widehat{\lambda}_{_{X\,\ell}}} - \frac{\sigma_{_{k\,\ell}}}{\lambda_{_{X\,\ell}}} \right)^2\,ds\,dt \right. \\
& + \int_{\mathcal{T} \times \mathcal{T}}  \left(\widehat{\Psi}_{_\ell}(s) -\Psi_{_\ell}(s)\right)^2\,\Phi_{_k}^2(t) \left(\frac{\sigma_{_{k\,\ell}}}{\lambda_{_{X\,\ell}}}\right)^2\,ds\,dt \\
&\left. + \int_{\mathcal{T} \times \mathcal{T}} \left(\widehat{\Phi}_{_k}(t) - \Phi_{_k}(t) \right)^2\,\Psi_{_\ell}^2(s) \,\left(\frac{\sigma_{_{k\,\ell}}}{\lambda_{_{X\,\ell}}}\right)^2\,ds\,dt \right]\,\left[1 + o_p(1)\right]\,.
\end{align*}
Since $\Psi_{_\ell}$ and $\Phi_{_k}$ are orthonormal, we conclude that  
\begin{equation}\label{f1CR}
f_{_1}\leq \sum\limits_{k = 1}^{J_{_1} - 1} \sum\limits_{\ell = 1}^{J_{_2} - 1} \left[\gamma_{_{k\,\ell}}+o_p(\gamma_{_{k\,\ell}})\right]\,,
\end{equation}
where
\begin{equation*}
\gamma_{_{k\,\ell}}=\left(\frac{\widehat{\sigma}_{_{k\,\ell}}}{\widehat{\lambda}_{_{X\,\ell}}} - \frac{\sigma_{_{k\,\ell}}}{\lambda_{_{X\,\ell}}}\right)^2 + \left( \frac{\sigma_{k\,\ell}}{\lambda_{_{X\,\ell}}} \right)^2\,\left(\Big\|\,\widehat\Psi_{_\ell} - \Psi_{_\ell}\,\Big\|_{_{L_2}}^2 + \Big\|\,\widehat{\Phi}_{_k}- \Phi_{_k}\,\Big\|_{_{L_2}}^2\right)\,.
\end{equation*}
Similarly, it can be shown that
\begin{equation*}
\Big|\,\widehat{\sigma}_{_{k\,\ell}} - \sigma_{_{k\,\ell}}\,\Big| \le C_1\,\left[\Big\|\widehat{\Psi}_{_\ell} - \Psi_{_\ell}\,\Big\|_{_{L_2}} + \Big\|\,\widehat{\Phi}_{_k} - \Phi_{_k}\Big\|_{_{L_2}} + \Big\|\,\widehat{C}_{_{X\,Y}} - C_{_{X\,Y}}\,\Big\|_{_{L_2}}\right]\,\left[1 + o_p(1)\right]\,.
\end{equation*}
Therefore Theorem \ref{T3} along with equation \eqref{PsiCR} imply that
\begin{equation}\label{PsiCR2}
\Big|\,\widehat{\sigma}_{_{k\,\ell}} - \sigma_{_{k\,\ell}}\,\Big|^2=O_p \left( \frac{1}{n}+\left[\frac{\log(n)}{m\,n} \right]^{\frac{2\,\nu}
{2\,\nu + 1}}  \right)\,.
\end{equation}
Now, equations \eqref{PsiCR}, \eqref{K,M}, \eqref{f1CR} and \eqref{PsiCR2} imply that
\begin{equation*}
f_{_1} = O_p\left(\left[\frac{1}{n} + \left(\frac{\log(n)}{m\,n} \right)^{\frac{2\,\nu}{2\,\nu+ 1}} \right]^{ \frac{\tau  + \gamma}{\tau  + \gamma  + 2}}\right).
\end{equation*}
By \eqref{sig}
\begin{align*}
  f_{_2}  &= \sum\limits_{k = J_{_1}}^\infty \sum\limits_{\ell = J_{_2}}^\infty \frac{\sigma_{_{k\,\ell}}^2}{\lambda_{_{X\,\ell}}^2} \\
 &\leq c_{_0}\,\left(\sum\limits_{k = J_{_1}}^\infty k^{- (\tau +1)} \right)\,\left( \sum\limits_{\ell = J_{_2}}^\infty \ell^{- (\gamma+ 1)} \right) \\
 &\leq c_{_0}\,J_{_1}^{- \tau} \,J_{_2}^{- \gamma }  \,.
\end{align*} 
Hence by \eqref{K,M}
\begin{equation*}
f_{_2} = O\left(\left[ \frac{1}{n} + \left(\frac{\log(n)}{m\,n}\right)^{\frac{2\,\nu}{2\,\nu + 1}} \right]^{\frac{\tau  + \gamma}{\tau + \gamma + 2}}\right)\, .
\end{equation*}
On the other hand, by the Cauchy Schwartz inequality
\begin{equation*}
f_{_3}^2\leq f_{_1} \,f_{_2}\,.
\end{equation*}
Thus
\begin{equation*}
f_{_3} = O_p\left(\left[ \frac{1}{n} + \left(\frac{\log(n)}{m\,n}\right)^{\frac{2\,\nu}{2\,\nu + 1}} \right]^{\frac{\tau  + \gamma}{\tau + \gamma + 2}}\right)\, ,
\end{equation*}
and finally we have 
$$\Big\|\,\widehat{\beta} -\beta \,\Big\|^2_{_{L_2}}=O_p\left(\left[ \frac{1}{n} + \left(\frac{\log(n)}{m\,n}\right)^{\frac{2\,\nu}{2\,\nu + 1}} \right]^{\frac{\tau  + \gamma}{\tau + \gamma + 2}}\right)\, .\qquad\blacksquare $$

\subsection{Proof of Theorem 4}
Let $X(\cdot)=\zeta \,\Psi(\cdot)$ and $Y(\cdot)=\xi\, \Phi(\cdot)$, where 
$Pr(\zeta =  - 1,\,\xi  =  - 1) = Pr(\zeta = 1,\,\xi  = 1) = \frac{1}{2}$.
Set $$ M_{_1} = c_{_{M_{_1}}}n^\frac{\tau +\gamma}{(\tau +\gamma +2)\,(2\,\nu -2\,\delta )}\,\quad \text{ for some constant }c_{_{M_{_1}}}>0\,.$$
For binary vectors $\bfmath{\omega},\,\bfmath{\omega}^\ast \in \Omega$, where 
$$\Omega =\lbrace\bfmath{\omega} =(\omega_{_1},\,\dots,\,\omega_{_{M_{_1}}})\,;\quad \omega_{_i}=-1\text{ or }+1 \rbrace= \lbrace \,\pm 1\,\rbrace^{M_{_1}}\,$$ 
define
\begin{equation*}
\Psi_{_\omega}(\cdot)=\left(\frac{M}{M_{_1}}\right)^{\frac{1}{2}} \sum\limits_{k = M_{_1}+1}^{2\,M_{_1}} \omega_{_{k - M_{_1}}}\, \left(\frac{k}{2}\right)^\delta \,\rho_{_k}^{\frac{1}{2}}\, \varphi_{_k}(\cdot) 
\end{equation*}
and
\begin{equation*}
\Phi_{_{\omega^\ast}}(\cdot) = \left(\frac{M}{M_{_1}}\right)^{\frac{1}{2}} \sum\limits_{k = M_{_1}+1}^{2\,M_{_1}} \omega^\ast_{_{k - M_{_1}}} \left(\frac{k}{2}\right)^\delta\, \rho_{_k}^{\frac{1}{2}}\, \varphi_{_k}(\cdot) \,,
\end{equation*} 
where $M$ is the bound given by the condition (a) in Section 4. Then
\begin{equation*}
\|\,\Psi_{_\omega}\,\|_{_{\mathcal{H}(K)}}^2 = \frac{M}{M_{_1}}\,\sum\limits_{k = {M_{_1}} + 1}^{2\,M_{_1}} \omega_{_{k - M_{_1}}}^2 \,\left(\frac{k}{2}\right)^{2\,\delta} \leq {M}
\end{equation*}
and
\begin{equation*}
\|\,\Phi_{_{\omega^\ast}}\,\|_{_{\mathcal{H}(K)}}^2 = \frac{M}{M_{_1}}\,\sum\limits_{k = M_{_1} + 1}^{2\,M_{_1}} \omega^{\ast\,2}_{_{k - M_{_1}}} \,\left(\frac{k}{2}\right)^{2\,\delta} \leq M\,.
\end{equation*}
Furthermore,
\begin{align*}
\|\,\Psi_{_\omega} - \Psi_{_{\omega^\ast}}\,\|_{_{L_2}}^2 &= \dfrac{M}{M_{_1}}\,\sum\limits_{k = M_{_1} +1}^{2\,M_{_1}} \rho_{_k} \,\left(\dfrac{k}{2}\right)^{2\,\delta}\,\left(\omega_{_{k - M_{_1}}} - \omega^\ast_{_{k - M_{_1}}}\right)^2\\
& \ge c_{_0}\,\left(\frac{M}{M_{_1}}\right)\,\sum\limits_{k = M_{_1} + 1}^{2\,M_{_1}} k^{- 2\,\nu +2\,\delta}\left(\omega_{_{k - M_{_1}}} -\omega^\ast_{_{k - M_{_1}}}\right)^2 \\
& \ge c_{_0}\,M\,M_{_1}^{- (2\,\nu  -2\,\delta + 1)}\,H(\bfmath{\omega},\,\bfmath{\omega}^\ast)\,,
\end{align*}
where $H(\bfmath{\omega},\,\bfmath{\omega}^\ast)$ is the Hamming distance between the vectors $\bfmath{\omega}$ and $\bfmath{\omega}^\ast$ .

\noindent
By Varshamov-Gilbert bound (see Tsybakov 2009) there exists a subset  $\lbrace \bfmath{\omega}^{(0)} ,\,\dots,\,\bfmath{\omega}^{(N)} \rbrace$ of $\Omega$ such that $\bfmath{\omega}^{(0)}=(0,\,\dots,\,0)$ and
\begin{equation*}
H(\bfmath{\omega}^{(j)},\,\bfmath{\omega}^{(k)}) \ge \frac{M_{_1}}{8}\,\quad \text{ for all } \quad 0 \le j < k \le N, \quad \text{ with }  
N \geq 2^{M_{_1}/8}\,.
\end{equation*}
Therefore
\begin{equation*}
\|\,\Psi_{\omega^{(j)}}  - \Psi_{\omega^{(k)}} \,\|_{_{L_2}}^2  \geq c_{_1}\, n^{-\frac{\tau + \gamma }{\tau + \gamma +2}}\,,
\end{equation*}
for some constant $c_{_1}>0$. Similarly, it can be seen that for $\lbrace \bfmath{\omega}^{(0)} ,\,\dots,\,\bfmath{\omega}^{(N)} \rbrace$ and some constant $c_{_2}>0$  
\begin{equation*}
\|\,\Phi_{\omega^{(j)}}  - \Phi_{\omega^{(k)}}\|_{_{L_2}}^2  \geq c_{_2} \,n^{-\frac{\tau + \gamma}{\tau + \gamma +2}}\,.
\end{equation*}
Let $\Pi_{_{1\,k}}$ be the probability measure of the triplet $(X,\,S,\,\varepsilon)$ such that $\varepsilon \sim N(0,\,\sigma_{_X}^2)$, $S$ follows a uniform distribution on $\mathcal{T}$, and
 $\Psi=\Psi_{_k} := (1 + \Psi_{\omega^{(k)}})/2$
 and suppose $\Pi_{_{1\,0}}$ be such that $\varepsilon \sim N(0,\,\sigma_{_X}^2)$, $S$ follows a uniform distribution on $\mathcal{T}$, and
 $\Psi=\Psi_0 : = 1/2$.
 Similarly define $\Pi_{_{2\,k}}$ and $\Pi_{_{2\,0}}$ for the triplet $(Y,\,T,\,\epsilon)$. Also define the product measures $\Pi_{_k}=\Pi_{1\,k}\otimes \Pi_{_{2\,k}}$ and $\Pi_{_0}=\Pi_{_{1\,0}}\otimes \Pi_{_{2\,0}}$. It is not hard to see that
\begin{equation*}
\|\,\beta_{_{k}} - \beta_{_{k'}}\,\|_{_{L_2 }}^2  = \|\,\Psi_{_k}  \otimes \Phi_{_k}  - \Psi_{_{k'}}  \otimes \Phi_{_{k'}}\,\|_{_{L_2}}^2  \geq c_{_3}\,n^{-\frac{\tau +\gamma}{\tau +\gamma +2}}
\end{equation*}
where $\beta_{_k}$ is the slope function when $\Psi=\Psi_{_k}$ and $\Phi =\Phi_{_k}$ and $c_{_3}>0$ is a constant. Conditional on $(S,\,\zeta)$, the random variables $U_{_{i\,j}}$ follow normal distribution with mean $X(S_{_{i\,j}})$ and variance $\sigma^2_{_X}$, respectively. Therefore, the Kullback-Leibler distance between probability measures $\Pi_{_{1\,k}}$ to $\Pi_{_{1\,0}}$ can be given by
\begin{align*}
\mathcal{K}(\,\Pi_{_{1\,k}}\,|\,\Pi _{_{1\,0}}\,)&=E_{_{1\,k}} \left( - \dfrac{1}{\sigma_{_X}^2}\,\sum\limits_{i = 1}^n \,\sum\limits_{j = 1}^{m_{_1}} \left[U_{_{i\,j}} - \zeta\, \Psi_{_0}(S_{_{i\,j}})\right]\,\left[\zeta\, \Psi_{_0}(S_{_{i\,j}}) - \zeta\,\Psi_{_k}(S_{_{i\,j}})\,\right]\right.\\
&\left. - \dfrac{1}{2\,\sigma_{_X}^2}\,\sum\limits_{i = 1}^n \sum\limits_{j = 1}^{m_{_1}} \,\zeta^2\,\left[\Psi_{_0}(S_{ij}) - \Psi_{_k}(S_{ij})\right]^2 \right)\,,
\end{align*}
where $E_{_{1\,k}}$ denotes the expectation with respect to the probability measure $\Pi_{_{1\,k}}$. Hence, we can bound $\mathcal{K}(\,\Pi_{_{1\,k}}\,|\,\Pi_{_{1\,0}}\,)$ by
\begin{align*}
\mathcal{K}(\,\Pi_{_{1\,k}}\,|\,{\Pi _{_{1\,0}}}) &= \frac{n\,m_{_1}}{2\,|\,\mathcal{T}\,|\,\sigma_{_X}^2}\,\|\Psi_{_k} - \Psi_{_0}\,\|_{_{L_2}}^2\\
& = \frac{n\,m_{_1}\,M}{2\,|\,\mathcal{T}\,|\,\sigma_{_X}^2\,M_{_1}}\,\sum\limits_{k = M_{_1} + 1}^{2\,M_{_1}} \,\rho_{_k}\,\left(\frac{k}{2}\right)^{2\,\delta}\,\left(\omega_{_{k - M_{_1}}} - \omega^\ast_{_{k - M_{_1}}}\right)^2 \\
& \le c_{_4}\,n\,m_{_1}\,\,M_{_1}^{- (2\,\nu  - 2\,\delta  + 1)}\,H(\,\bfmath{\omega},\,\bfmath{\omega}^\ast)\\
& \le c_{_4}\,n\,m_{_1}\,M_{_1}^{- (2\,\nu  - 2\,\delta )}\,,
\end{align*}
for some constant $c_{_4}>0$ . 
Similarly, it can be shown that, for some constant $c_{_5}>0$,
$$
\mathcal{K}(\,\Pi_{_{2\,k}}\,|\,\Pi_{_{2\,0}}\,) \le c_{_5}\,n\,m_{_2}\,M_{_1}^{-(2\,\nu  - 2\,\delta)}\,.
$$
Thus
\begin{align*}
\mathcal{K}(\,\Pi_{_k}\,|\,\Pi_{_0}) & = \mathcal{K}(\,\Pi_{_{1\,k}}\,|\,\Pi_{_{1\,0}}\,) + \mathcal{K}(\,\Pi_{_{2\,k}}\,|\,\Pi_{_{2\,0}}\,) \\
& \le c_{_6}\,n\,(m_{_1} + m_{_2})\,M_{_1}^{- (2\,\nu  - 2\,\delta)}\,,
\end{align*}
for some constant $c_{_6}>0$ .
By using Fano's Lemma (see Tsybakov 2009) and taking $c_{_{M_1}}$ large enough, we get
\begin{align*}
\max\limits_{0 \le j \le N} E_{_j}\left[\,\Big\|\,\widetilde{\beta} - \beta_{_j}\Big\|_{_{L_2}}^2\,\right] \ge \frac{c_{_3}}{2}\,n^{- \frac{\tau  + \gamma}{\tau + \gamma + 2}}\,\left(1 - \frac{c_{_6}\,n\,(m_{_1} + m_{_2})\,M_{_1}^{- (2\,\nu  - 2\,\delta )} + \log(2)}{\log(N)} \right) \asymp n^{- \frac{\tau  + \gamma }{\tau  + \gamma  + 2}}\,,
\end{align*}
where $E_{_j}$ denotes the expectation with respect to the probability measure $\Pi_{_j}$. So,
$$\lim\sup\limits_{n \to \infty}\inf\limits_{\widetilde{\beta}} \sup\limits_{F \in \mathcal{F}_{_0} (\nu ;\,M,\,c,\,c_{_0})} P_F\left( \Big\|\,\widetilde{\beta}  - \beta\, \Big\|_{_{L_2}}^2 > d\,n^{- \frac{\tau  + \gamma }{\tau + \gamma  + 2}} \right) > 0\,.\quad \blacksquare$$
\section*{Acknowledgement}
The research of Dr. Chenouri was partially supported by the Natural Science and Engineering Research Council of Canada.

\end{document}